%% file: Planck_XVI.tex
\documentclass[traditabstract,longauth]{aa}
\usepackage[nonamebreak]{natbib}
\usepackage[stable]{footmisc}
\usepackage{txfonts}
\usepackage{xspace}
\usepackage{placeins}
\usepackage{natbib}
\bibpunct{(}{)}{;}{a}{}{,} 
\usepackage{graphicx}
\usepackage{epstopdf}
\usepackage{ifthen}
\usepackage{microtype}
\usepackage[breaklinks, colorlinks, citecolor=blue]{hyperref}
\usepackage{subfigure}
\usepackage[table,usenames,dvipsnames]{xcolor}

\usepackage{dblfloatfix}
\usepackage{float}

\input macros.tex

\input Planck.tex

\newcommand{\wwwminuit}{\url{http://seal.web.cern.ch/seal/work-packages/mathlibs/minuit/index.html}}
\newcommand{\wwwpla}{\url{http://www.sciops.esa.int/index.php?project=planck&page=Planck_Legacy_Archive}}
\newcommand{\wwwcosmomc}{\url{http://cosmologist.info/cosmomc/readme_planck.html}}
\newcommand{\wwwbobyqua}{\url{http://www.damtp.cam.ac.uk/user/na/NA_papers/NA2009_06.pdf}}
\newcommand{\wwwjames}{\url{http://seal.cern.ch/documents/minuit/mnerror.pdf}}
\newcommand{\citeg}[1]{\citep[e.g.,][]{#1}}

\begin{document}
\input title
\input{P11a_Parameters_Frequentist_Estimation_authors_and_institutes.tex}

\input abstract

\date{\vspace{-0.2in} \today}
\titlerunning{Cosmological parameters}
\maketitle


\section{Introduction}

This paper, one of a set associated with the 2013 release of data from
the \Planck\footnote{\Planck\ (\url{http://www.esa.int/Planck}) is a
  project of the European Space Agency (ESA) with instruments provided
  by two scientific consortia funded by ESA member states (in
  particular the lead countries France and Italy), with contributions
  from NASA (USA) and telescope reflectors provided by a collaboration
  between ESA and a scientific consortium led and funded by Denmark.}
mission \citep{planck2013-p01}, describes a frequentist estimation of cosmological parameters using profile likelihoods.  

Parameter estimation in cosmology is predominantly performed using Bayesian inference, particularly following the introduction of 
Markov chain Monte Carlo (MCMC) techniques \citep{christensen}. Many scientists in
the field use the sophisticated \cosmomc\footnote{Available from \wwwcosmomc} software \citep{cosmomc} to
study cosmological parameters, and several experiments provide 
ready-to-use plugins for it.
The \planck\ satellite  mission has recently released
high-quality data on the cosmic 
microwave background (CMB) temperature anisotropies.\footnote{Available from
  \wwwpla} The analysis of
the cosmological parameters \citep{cosmo} is based on
Bayesian inference using a dedicated version of \cosmomc.

In this methodology, the likelihood leads to the posterior
distribution of the parameters once it has been multiplied by some
prior distribution that encompasses our knowledge before the
measurement is performed. For \planck, wide bounds on uniform
distributions have typically been used. However the choice of a particular set of parameters
for MCMC sampling, such as the efficient ``physical basis'' \citep{basephys} used in \cosmomc, may
also be viewed as an implicit prior choice. 

Frequentist methods do not need priors, other than
that some limits on the explored domain are used in practice and can
be seen as the bounds of some ``uniform priors''. 
The maximum likelihood estimate (MLE) does not depend on the
choice of the set of parameters, since it possesses the property of
\textit{invariance}: if $\hat \theta$ represents the MLE of the parameter
$\theta$, then the MLE of any function $\tau(\theta)$ is $\hat
\tau=\tau(\hat \theta)$. 
This means that one can compute the MLE with any set of parameters.
As we will see in \sect{sec:stat}, this property is
powerful and can be used to obtain asymmetric confidence intervals.

The multi-dimensional solution is only one aspect of parameter
estimation and we are also interested in statements on individual parameters. 
In the MCMC procedure, once the chains have converged
this is obtained through \textit{marginalization}, which is performed
by a simple 
projection of the samples onto one or sometimes two axes.
This may however lead to so-called ``volume effects'', where the mean of the
projected distribution can become incompatible with the
multi-dimensional MLE \citeg{hamman1}.
In the frequentist framework, one instead builds \textit{profile likelihoods}
\citep{proflkl} for individual variables and, by construction, the
individual parameter estimates match (up to numerical accuracy) the MLE values.

Such a method has already been used by \citet{yeche} with Wilkinson Microwave Anisotropy Probe (\wmap) 
data for a nine-parameter fit. The high
sensitivity of data from \planck\ and from the ground-based South Pole Telescope (SPT) and Atacama Cosmology Telescope (ACT) projects requires the simultaneous fit of a larger
number of parameters, up to about $40$, with some nuisance ones being poorly constrained.
We therefore need to precisely tune a high-quality minimizer, as will be described in this paper.

MCMC sampling is sometimes used to perform a ``poor-man's'' determination
of the maximum likelihood \citeg{reid}: one bins a given
parameter and reports the sample of maximum likelihood in other dimensions. As
pointed out in \citet{hamann}, in many dimensions it is most likely
that the real maximum was never reached in any reasonably-sized chain. The authors suggest
changing the temperature of the chain, but this still requires running
lengthy evaluations of the likelihood and is less straightforward than
directly using a multi-dimensional minimization algorithm.

In this article, we investigate whether the use of
priors or marginalization can affect the determination of the
cosmological parameters by comparing the published Bayesian results to a frequentist method.
For the base \lcdm\ model, it happens that the cosmological parameter posteriors are
essentially Gaussian, so it is expected that frequentist and Bayesian
methods will lead to similar results. In extensions to the standard
\lcdm\ model this is however not true for some parameters (e.g., the sum
of neutrino masses), and priors have been shown to play some 
role in parameter determination \citep{hamman1,gonzalez,hamann}. 
Given the sensitivity of the \planck\ data, statistical methodologies may
matter, and this issue is scrutinized in this work.

In order to build precise profile likelihoods in a high-dimensional
space (up to about 40 dimensions), we need a powerful minimizer. We use the mature and widely-used \minuit software \citep{minuit}. We interfaced it to the modular \class
Boltzmann solver \citep{classII} which, from a set of input cosmological
parameters, computes the corresponding temperature and
polarization power spectra that are tested against the \planck\ likelihood.
This required that we tune the \class precision parameters to a level where the numerical
noise can be handled by our minimizer, as is described in
\sect{sec:class}. In \sect{sec:minuit}, we describe our \minuit minimization
strategy, and cover in \sect{sec:stat} the
basics of the frequentist methodology to estimate unknown parameters
based on the properties of profile likelihoods.
The data sets we use are then discussed in \sect{sec:datasets}.
We give results for the \lcdm\ parameters in \sect{sec:lcdm} and finally 
investigate, in \sect{sec:neutrinos}, a case where the posterior distribution is far
from Gaussian, namely  the neutrino mass
case. Additionally, the Appendix gathers
some comments on the overall computation time of the method.

\section{Method}
\label{sec:method}

\subsection{The Boltzmann solver: \class}
\label{sec:class}

To compute the relevant CMB power spectra from a cosmological model, we
need a ``Boltzmann solver'' that numerically evolves the
coupled perturbation equations in an expanding universe.  
While \camb is used in the \cosmomc sampler, we prefer
to use the \class (v1.6) software \citep{classII}.
It offers a rigorous way to control the accuracy of
output quantities through a comprehensive list of precision
parameters \citep{classI}. 
While one can use some high-speed/low-quality settings
to perform MCMC sampling because the random nature of the algorithm smooths out discontinuities,  
this is no longer the case here when searching for an extremum, which
requires precise computation of numerical derivatives. 
Equally, due to computation time, one cannot use 
precision settings that are too extreme, so a trade-off with \minuit convergence
has to be found.

As we will see in \sect{sec:stat}, 68\% confidence intervals are obtained by
cutting $\chi^2\equiv-2\ln {\cal L}$ values at one. We therefore 
need the numerical noise to be much less than unity.

\begin{figure}[h!]
  \centering
  \includegraphics[width=88mm,angle=0]{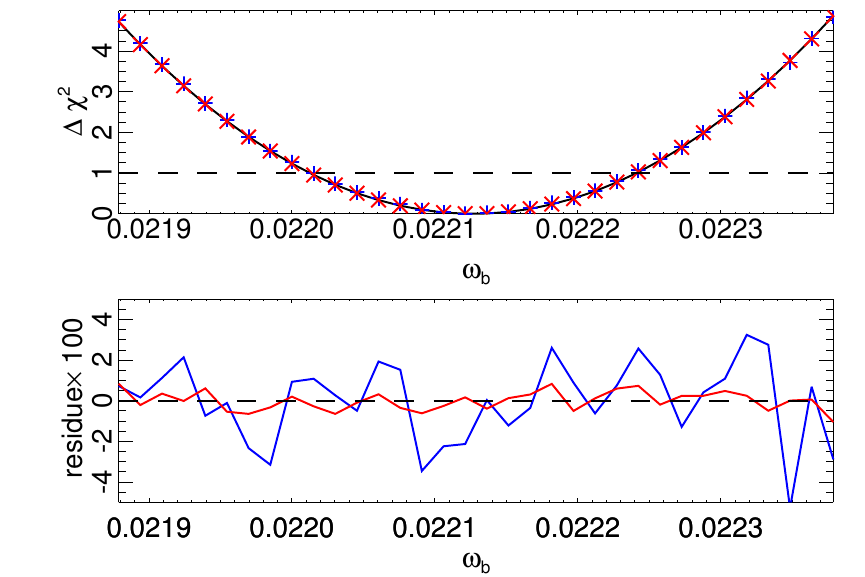}
  \caption{\textit{Upper panel}: the $\omb$ parameter is scanned (keeping all other
    parameters fixed to their best-fit values) and the \planck\
    $\chi^2$ values are shown on the vertical axis. Blue points are
    obtained with the \class 
    default settings, and red
    ones with our high-precision ones. A smooth parabola is fit and shown
  in black. \textit{Lower panel}: residuals with respect to the parabola. The r.m.s.\ of this noise is improved from $0.02$ for the default
  settings to $0.005$ for the high-precision ones.}
  \label{fig:scan}
\end{figure}

Starting from the \planck\ likelihood code, described in \sect{sec:datasets},
we fix all parameters to their published best-fit
values \citep{cosmo} and scan a given parameter $\theta$.
We compute the $\chi^2(\theta)$ curves and subtract a smooth component
to estimate the amplitude of the numerical
noise. According to the precision settings, trade-off between the
amplitude of this noise and the computation time can then be found.
An example with two precision settings is shown in Fig.~\ref{fig:scan}
for $\theta=\omb=\Omb h^2$ which is used as our benchmark.

We have determined a set of high-precision
settings which achieves sufficient smoothness of the \planck\
likelihood for the fits to converge, with an increase of only about a
factor two in the code computation speed \wrt the default
``fast'' settings. The values of the settings are reported in Table~\ref{tab:prec}.

\begin{table}[h!]
\begin{center}
\input Tables/CLASS_setup.tex
\caption{\label{tab:prec}
Values of the non-default precision parameters for \class used
  for the \minuit minimization.}
\end{center}
\end{table}

We also found that working with the Thomson scattering optical depth
$\tau$ is numerically less stable 
than using the reionization redshift \zreio, which defines where the
reionization fraction is half of its maximum. We therefore use \zreio as a primary parameter. The relation
to $\tau$, for a $\tanh$-based ionization profile and a fixed
$\Delta \zreio=0.5$ width, is given in \citet{lewis2008}.

Since we will compare our results to the previously-published ones, we need to ensure our \class
configuration 
reproduces the \camb-based results of \citet{planck2013-p11}.
For this purpose, we use the \planck\ \lcdm\ best-fit solution and
compute its $\chi^2$ value and compare with the published results in
Table~\ref{tab:CLASSvsCAMB}.
The agreement is good.
The slight discrepancy is typical of the differences between \class and \camb implementations
\citep{classIII}, so we consider our setup to be properly
calibrated. From now on, we perform consistent comparisons using only \class.

\begin{table}[t]
\begin{center}
\input Tables/CLASSvsCAMB.tex
\caption{\label{tab:CLASSvsCAMB}
Comparison of the $\chi^2$ values of the \planck\ best-fit solution from
  \citet{planck2013-p11}, based on \camb, to our \class-based
  implementation, for the \cmb and \cmbao data sets.}
\end{center}
\end{table}

\subsection{Minimizing  with \minuit}
\label{sec:minuit}


We chose to  work with the powerful \minuit
package \citep{minuit}, a well-known 
minimizer originally developed for high-energy physics and used
recently for the Higgs
mass determination with a simultaneous fit of 354 parameters \citep{atlas}.
While its roots trace back to the 1970s, it has been
continually improved and rewritten in \texttt{C++} as
\texttt{Minuit2}, 
which is the version we use. \minuit is a toolbox including several
algorithms that can be deployed depending on the problem under
consideration. We refer the reader to the user
guide\footnote{Available from \wwwminuit} for a detailed description
of the procedures we used.

For cosmological parameter estimation with the \planck\ data, we
executed the following strategy
\begin{enumerate}
\item Starting from the \planckcoll\ published values and using the
  high-precision \class settings described in \sect{sec:class},
  we minimize the $\chi^2$ function 
  using the \texttt{MIGRAD} algorithm, which is based on Fletcher's switching algorithm \citep{fletcher}. 
  All parameters are bounded by large (or physical) limits during this exploration. 
\item Once a minimum is found, we release all cosmological parameter
  limits and again perform the \texttt{MIGRAD} minimization. The limits on
  nuisance parameters are kept in order to avoid exploring unphysical regions.
\item Finally, we use the \texttt{HESSIAN} procedure which refines the local covariance matrix.
\end{enumerate}

\texttt{MIGRAD} belongs to the category of
  \textit{variable metric methods} \citeg{vmm} which build the ``expected
 distance to minimum'' (EDM) that represents (twice) the vertical
 distance  to the $\chi^2$ minimum if the function is truly quadratic and the
  gradient exactly known. It can serve as a figure of merit for the 
  convergence and will be used to reject poor fits.

The outcome of this procedure is the minimum $\chi^2$
solution together with its Hessian matrix. This solution represents
the MLE, but, since the problem is highly non-linear (in particular in $H_0$), the
Hessian is only a crude approximation to the parameter
uncertainties.\footnote{As discussed in \wwwjames} The
complete treatment is through the construction of profile likelihoods.

\subsection{Profile likelihoods}
\label{sec:stat}

The MLE (or ``best-fit'' or $\chi^2_\mathrm{min}$) is the global maximum likelihood
estimate given the entire set of parameters (cosmological and nuisance).
One can choose to isolate one parameter (hereafter called $\theta$)
and for fixed values of it look for the maximum of the likelihood function in all other dimensions.
One \textit{scans} $\theta$ within some range and, for each fixed 
value, runs a minimization \wrt all the other parameters. The minimum
$\chi^2$ value is reported for this parameter $\theta$, which allows one to
build the profile likelihood $\chi^2(\theta)$.
The procedure ensures the minimum of $\chi^2(\theta)$ appears at
the same value as the MLE, avoiding the potential volume
effects mentioned in the introduction.

A confidence region, which has the correct frequentist coverage
properties, can then be extracted from the likelihood ratio
statistic, or equivalently the $\dchi(\theta)=\chi^2(\theta)-\chi^2_\mathrm{min}$ distribution.
For a parabolic $\chi^2(\theta)$ shape (\ie Gaussian estimator
distribution), a $1-\alpha$ level confidence interval is obtained by
the set of values $\dchi(\theta) \le \chi^2_1(\alpha)$, where
$\chi^2_1(\alpha)$ denotes the $1-\alpha$ quantile of the chi-square
distribution with one degree of freedom, and is $1,~2.7$, and $3.84$ for
$1-\alpha=68,~90$ and $95\%$ respectively \citeg{jamesbook}.

It is less well known that if the profile likelihood is non-parabolic, one can \textit{still}
build an approximate confidence interval using the same recipe, because the full likelihood ratio
has the \textit{invariance} property mentioned in
the introduction: one can estimate any monotonic function of $\theta$
and make the same inference not only on the MLE but on \textit{any likelihood
ratio}. For example, one can build the $\dchi(\As)$ distribution from the $\dchi(\lnAs))$
profile by simply switching the $\lnAs\to\As$ axis.
Formally, when the profile likelihood is non-parabolic, one can still \textit{imagine}
a transformation that would make it quadratic in the new variable. One
would then apply the parabolic cuts described previously and, by the
invariance property, the same inference on the original variable would
be obtained. Therefore we can find an (asymmetric) confidence
interval by cutting the non-parabolic \dchi curve at the same
$\chi^2(\alpha)$ values. This method, sometimes called \texttt{MINOS} 
(the name of the routine that first implemented it in
\minuit), is long known in the statistics field \citep{proflkl}.
It is exact up to order ${\cal O}(1/N)$ \citep{jamesbook}, $N$ being
the number of samples, and is in practice excellent unless $N$ is very
small.

Nevertheless, the profile-likelihood-based confidence intervals must be revisited in the case where
the estimate lies near a physical boundary. This will be
performed in \sect{sec:neutrinos} for the neutrino mass case.

\section{Data sets}
\label{sec:datasets}

As our purpose is to compare the frequentist methodology
to the Bayesian one, we focus on exactly the same data and parameters
as in \citet{planck2013-p11} and refer the reader to \citet{planck2013-p08}
for their exact definitions.
Since the CMB, baryon acoustic oscillation (BAO), and CMB-lensing data were found to be in excellent
agreement, we will consider the following likelihood combinations.

The \cmb data set consists of the following likelihoods: 
\begin{itemize}
\item the \planck\ 2013 data in both low and high $\ell$ ranges;
\item the \wmap low-$\ell$ polarization data (referred to as \WP\ in
  the \planck\ paper);
\item the SPT \citep{R12}+ACT \citep{D13} high-$\ell$ data, referred to as highL.
\end{itemize}
The combined likelihood, obtained by multiplying the three, includes 31 nuisance parameters, related to
the characterization of the unresolved foregrounds, the effective beam, 
and to the inter-calibration of the \planck\ and highL power spectra.

The \rm{BAO} data set consists of a Gaussian likelihood based on the scale
measurements from the \textit{6dF} \citep{6dF}, SDSS \citep{SDSS}, and BOSS \citep{BOSS} experiments, combined as in \citet{planck2013-p11}.

For the neutrino mass case, we will also use the \planck\
lensing likelihood \citep{planck2013-p12}, based on the measurement of the deflection power spectrum.

\begin{table*}[htbp]
\begin{center}
\input Tables/LCDM_bestfit.tex
\caption{\label{tab:bf} Best-fit comparison. Values of all parameters at the minimum of the $\chi^2$ function as determined by \cosmomc
   in \citet{planck2013-p11} and by the \minuit implementation described here, for the
   \cmb and \cmbao data sets. The first six parameters define the \lcdm\ cosmology. 
 The last line shows the $\chi^2$ value at the minimum.
}
\end{center}
\end{table*}

\section{Results}
\label{sec:results}
\subsection{The base \lcdm\ model}
\label{sec:lcdm}

We begin by revisiting the global best-fit solution (MLE) using this new minimizer,
over all 37 parameters, on the \cmb and \cmbao data sets.
We use the Hubble constant ($H_0$) instead of the CMB acoustic
scale ($\theta_\mathrm{MC}$), which is not available within \class, and \zreio instead
of $\tau$ since it is more stable as discussed in \sect{sec:class}.
The new minimum is given in Table~\ref{tab:bf} and compared to the
results previously released in \citet{planck2013-p11} 
, which were obtained with another minimizer.\footnote{Named
  \texttt{BOBYQUA} and described in \wwwbobyqua} In both cases we find a
slightly lower $\chi^2$. 
On the cosmological side we find very similar parameters, except for $\zreio$ which is slightly shifted.
On the nuisance parameters side, results are also similar, but we are
now  sensitive to the SZ--CIB cross-correlation parameter $\xi^\mathrm{tSZ-CIB}$
while the \planckcoll\ minimum was not shifted from its zero initial value.
Additionally the estimated kinetic SZ amplitude $A^\mathrm{kSZ}$ is more stable when including the BAO data set. 

\begin{figure*}[h]
\centering
  \begin{tabular}{cc}
\includegraphics[width=.49\textwidth,angle=00,clip]{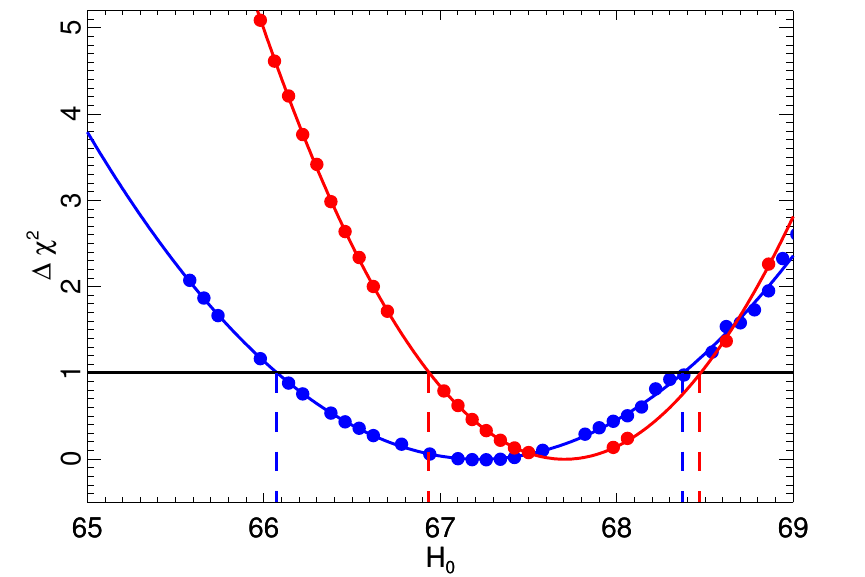}&
\includegraphics[width=.49\textwidth,angle=0]{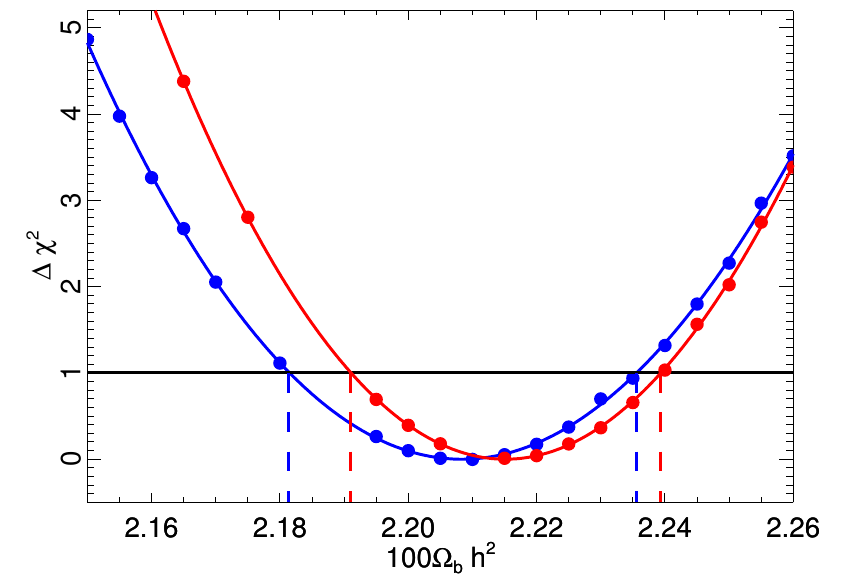}\\
\includegraphics[width=.49\textwidth,angle=0]{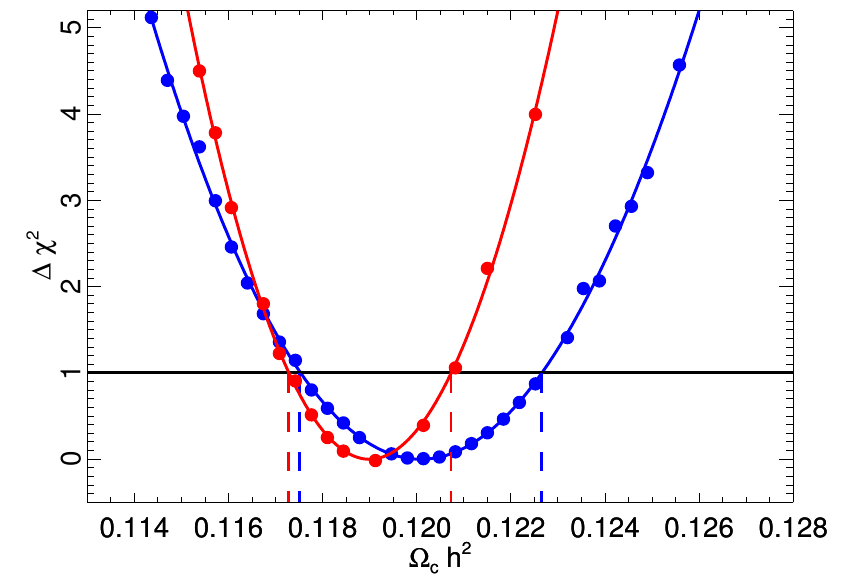} &
\includegraphics[width=.49\textwidth,angle=0]{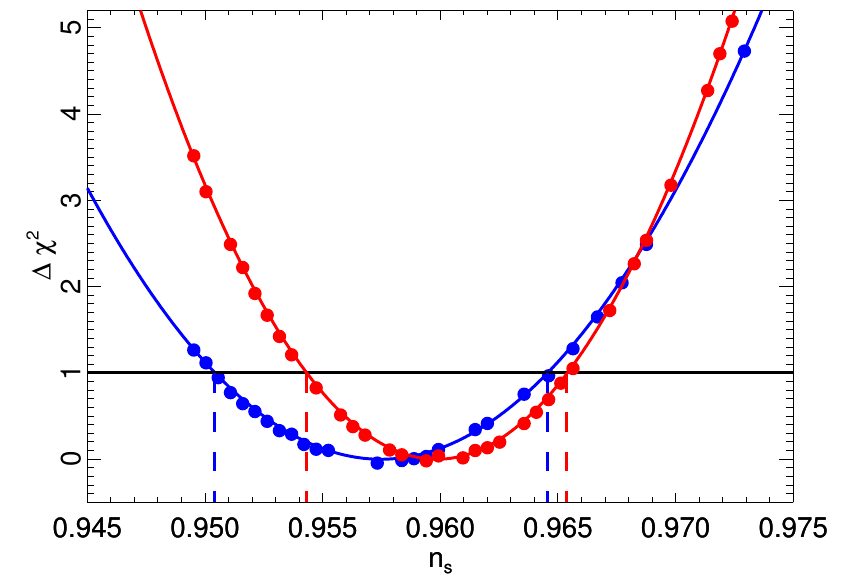}\\
\includegraphics[width=.49\textwidth,angle=0]{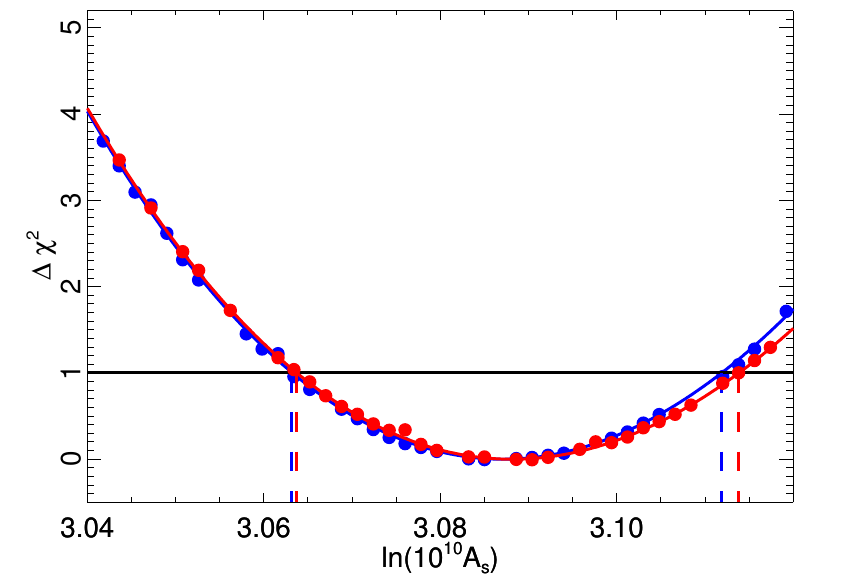}&
\includegraphics[width=.49\textwidth,angle=0]{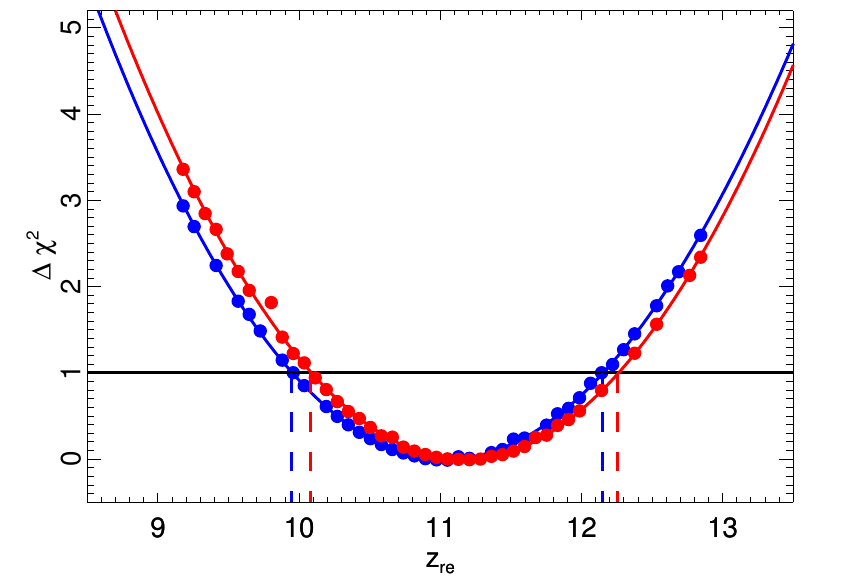} 
  \end{tabular}
\caption{Profile likelihoods  (\dchi) reconstructed for each \lcdm\ cosmological parameter,
from the \cmb (blue) and \cmbao (red) data sets. Each point
is the result of a 36-parameter minimization. We reject the points
that are outliers of the expected distance to minimum (EDM,
\sect{sec:minuit}) distribution. Curves
are fits to a third-order polynomial. 68\% confidence
intervals are obtained by thresholding these curves at unity, and their
projections onto the parameter axis are shown.}
\label{fig:proflcdm}
\end{figure*}

We then build the profile likelihoods by scanning each cosmological
parameter and computing the $\chi^2$  minimum in the remaining 36
dimensions at each point.
Fig.~\ref{fig:proflcdm} shows the reconstructed profiles.
They are found to be mostly parabolic, but we still fit them with a
third-order polynomial in order to measure any deviation from a symmetric error, and threshold
them at unity in order to obtain the 68\% frequentist confidence level
interval as explained in \sect{sec:stat}.
Results are reported in Table~\ref{tab:proflcdm} and compared there to the \planckcoll\ posterior distributions.
In most cases the values and errors we obtain are in good agreement with
the Bayesian posteriors, demonstrating that the \planckcoll\ results, 
for the \lcdm\ model, are not biased by a particular choice of parameters (implicit priors)
or by the marginalization process (volume effects).

\begin{table*}[t]
\begin{center}
\input Tables/LCDM_profile.tex
\caption{\label{tab:proflcdm}
Results of the profile-likelihood analysis (\ie, this work)
  for the cosmological parameters, using the \cmb and \cmbao
  data sets. They are compared to the \planck\
  MCMC posterior results taken from Table 5 of \citet{planck2013-p11}.}
\end{center}
\end{table*}

By comparing the mean values of Table~\ref{tab:proflcdm} to the
best-fit ones (Table~\ref{tab:bf}) we observe that the minima coincide at
the percent level, as expected for this frequentist method.

Since we observe some difference in the reionization parameter \zreio, we also
perform the profile-likelihood analysis with the \planck\ data alone
and obtain
\begin{equation}
  \zreio = 13.3^{+2.8}_{-3.3} \qquad (\plancko,\text{profile~likelihood}),
\end{equation}
while the \planckcoll\ reports
\begin{equation}
  \zreio = 11.4^{+4.0}_{-2.8} \qquad (\plancko,\text{MCMC~posterior}).
\end{equation}
The results, \textit{using exactly the same data}, are different. We
believe that these new results are robust since the profile-likelihood method is particularly well suited for this
case. Indeed \zreio is \textit{fixed} in each step so that the minimization does not suffer
from the classical $(\As,\zreio$) degeneracy due to the normalization
of the temperature-only power spectrum.
In contrast, the MCMC method relies strongly on the priors used on both $\As$ and $\zreio$.
We find that it is the inclusion of the \wmap polarization data
that pulls down this value to $\zreio=11.0\pm1.1$, as reported in Table~\ref{tab:proflcdm}.

\subsection{Mass of standard neutrinos}
\label{sec:neutrinos}

Since the cosmological parameter posterior distributions for the \lcdm\ model are mostly Gaussian
(parabolic $\chi^2$), the Bayesian and frequentist approaches lead to similar results. However, we may expect greater differences
when including neutrino masses in the model, for which the
marginalized posterior distribution is peaked towards zero.

CMB measurements are sensitive to the sum of neutrino mass eigenstates
$\mnu$  through several effects reviewed in detail in
\citet{lesgourgues}. For large values \mbox{$\mnu \gtrsim
1.3\ev$}, the neutrinos' non-relativistic transition happens before
decoupling and the integrated Sachs--Wolfe effect 
reduces the amplitude of the first acoustic peak. For
lower mass values, neutrino free-streaming erases small-scale matter
fluctuations and accordingly reduces the CMB lensing power. This
in turn affects the lensed \cl spectrum, especially its high-$\ell$ 
part, and explains the gain when including \spt data.
Furthermore, since according to oscillation experiments at least two
neutrinos are non-relativistic today \citep{pdg}, 
the matter--radiation equality scale factor, which is strongly constrained by the
\planck\ data, reads:
\begin{equation}
  \frac{a_\mathrm{eq}}{a_0}=\frac{\omega_\mathrm{r}}{\omega_\mathrm{m}-\omega_\nu},
\end{equation}
where $\omega_\mathrm{r}$, $\omega_\mathrm{m}$, and $\omega_\nu$ are the physical densities of
radiation, matter, and massive neutrinos respectively, \ie
$ \omega_\mathrm{r}=\Omega_{\mathrm r} h^2$, \mbox{$\omega_\mathrm{m}=\Omega_\mathrm{m} h^2$}, and $\omega_\nu=\Omega_\nu
  h^2 \simeq 10^{-3} \mnu/0.1\ev$.
The quantities $\omega_\nu$ and $\omega_\mathrm{m}$ are clearly degenerate, and so
any data set that helps in reducing the CMB geometrical degeneracies by providing a measurement at
another scale indirectly benefits $\mnu$.
Robust observables, compatible with the \planck\ \lcdm\ cosmology, are
the BAO scale measurement around $z\simeq 0.5$ and/or the CMB-lensing trispectrum that probes
matter structures around $z\simeq 2$.

An unexpected result found by \citet{planck2013-p11} is that the 95\% confidence upper limit on
$\mnu$ obtained from \planck\ data is \textit{worsened} when including the
lensing trispectrum information (the 95\% upper limit goes from 0.66 to 0.84). How can
the addition of new information weaken the limit? Is this an effect of the
Bayesian methodology, which computes \textit{credible} intervals and
where such effects may arise when combining incompatible data?
Naively, in a frequentist analysis adding some information \citep[in the Fisher sense,
see][]{jamesbook} can only lower the size of
confidence interval, since the profile-likelihood ``error'' (its
curvature at the minimum) can only \textit{decrease} 
and thresholding it at a constant value should only lead to a smaller region.

\begin{figure*}[tbp]
  \centering
\includegraphics[width=120mm]{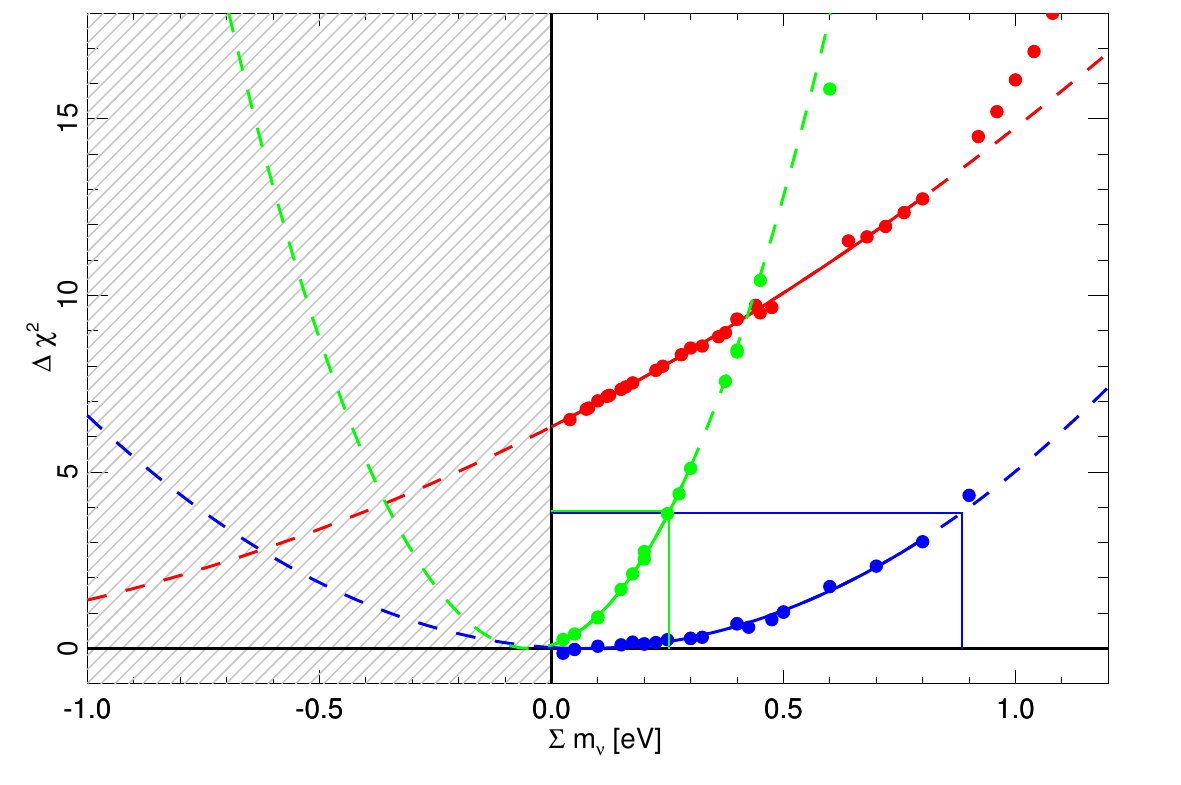}
\caption{Neutrino mass profile likelihood for the \cmb
(red), CMB+lensing (blue), and CMB+lensing+BAO (green) data sets. Each point is the result of a
37-parameter fit which can only be computed in the positive region. The
points are fit by a parabola and extrapolated into the
negative region. For the \cmb only case, the parabolic fit agreement is
poor and is only shown for discussion.
The coloured green/blue lines are used to set 95\% confidence upper limits according to the
Feldman--Cousins prescription, as described in the text.}
\label{fig:profmnu} 
\end{figure*}

We construct the profile likelihood for $\mnu$. It is shown in Fig.~\ref{fig:profmnu}
for the \cmb, CMB+lensing and CMB+lensing+BAO data sets.
We observe an intriguing feature with the \cmb data set.
Even though the parabolic fit of the profile likelihood is
poor, the minimum lies
at about $-2.5\sigma$ into the unphysical negative region.
When adding the lensing trispectrum information, it shifts back to a value compatible with zero.
We do not yet have a proper understanding of why this is happening, but note a possible connection
to the \Al\ issue discussed in \citet{planck2013-p11}, where this
phenomenological parameter is discrepant from unity by about $2\sigma$ using the \cmb
data set, but lowered to $1\sigma$ when adding the lensing information.

We can then understand why our previous argument on reducing the confidence interval by adding information is invalid
\textit{near a physical boundary}, even in a frequentist sense. 
If we consider a constant threshold of the profile likelihood (for instance around 8 on
Fig. \ref{fig:profmnu}) we may end up with an upper limit that is
\textit{smaller} (even though the curvature is larger) when omitting
the lensing information, because of the shift of the minimum into the
unphysical region. This resembles the Bayesian result.

However the methodologies shows their differences in this situation. 
In the Bayesian case, when combining somewhat incompatible data sets 
within a model the credible region enlarges to account for it. 
In the frequentist case, thresholding the profile likelihood is incorrect
and we apply instead the \citet{FC} prescription. Within this
classical framework, there is a decoupling of the confidence level of the
goodness of fit probability from the one used in building the
confidence interval. Unlike in the Bayesian case, one first tests the
consistency of the data with the model, and then constructs the
confidence interval (at some given level) only for the candidates
that fulfil it. 
In our case, a minimum at $-2.5\sigma$ is very unlikely (below 1\%
probability) and we will therefore not consider it in the following.

We give in Table~\ref{tab:fitparam} the
parameters of the parabolic fits
$\chi^2(\mnu)=\chi^2_\mathrm{min}+\left[ (\mnu-m_0)/\sigma_\nu\right]^2$.
We only use points within $2\sigma_\nu$ from
zero, since the function is not necessarily quadratic far from its
minimum. In the following we will vary this cut. We report here the
numbers that lead to the largest final limit.

\begin{table}[h]
\begin{center}
\input Tables/MNU_fit.tex
\caption{\label{tab:fitparam}
Estimates of the minima positions ($m_0$) and curvature
  ($\sigma_\nu$) from the parabolic fits of Fig.~\ref{fig:profmnu} for the
data sets including lensing. The range of points used corresponds roughly to $2\sigma$.}
\end{center}
\end{table}

The classical Neyman construction of a confidence interval has some
inherent degree of freedom in it \citeg{pdg}. The Feldman--Cousins prescription, that
is most powerfull near a physical a boundary, is to introduce an
ordering based upon the likelihood ratios $R$:

\begin{equation}
  \label{eq:rfc}
  R=\frac{{\cal L}(x |\mu)}{{\cal L}(x |\mu_\mathrm{best})} \,,
\end{equation}
where $x$ is the measured value of the sum of neutrino masses $\mnu$, $\mu$ is the true
value, and $\mu_\mathrm{best}$ is the best-fit value of $\mnu$, given the data and
the physically-allowed region for $\mu$. Hence
we have $\mu_\mathrm{best}=x$ if $x\ge 0$, but $\mu_\mathrm{best}=0$ if $x< 0$, and 
the ratio $R$ is given by \citep{FC}:
\begin{eqnarray}
  \label{eq:rfc2}
\exp({-(x-\mu)^2/2}) & \mathrm{for} \; x>  0;\\
\exp({x\mu-{\mu^2/2}}) & \mathrm{for} \; x \le 0.
\end{eqnarray}
We then search for an interval
$[x_1,x_2]$ such that $R(x_1)=R(x_2)$ and
\begin{equation}
  \int_{x_1}^{x_2}{\cal L}(x|\mu)dx=\alpha,
\end{equation}
with $\alpha=0.95$ as the confidence level. These intervals are tabulated in \citet{FC}.

We obtain  
the confidence interval $[\mu_1,\mu_2]$ for each $x=m_0/\sigma_\nu$ extracted from
the parabolic fit to the $\chi^2$ profile as given in Table~\ref{tab:fitparam}.
The upper limits are then simply $\mu_2\times \sigma_\nu$. 

We give our final results in Table~\ref{tab:fit} and compare them to
the \planck\ Bayesian ones of \citet{planck2013-p11}. The agreement is impressive, despite the
use of two very different statistical techniques.
Finally, we varied the range of points used in the parabolic fit and the
limits we obtain are always lower than the one reported in
Table~\ref{tab:fit}, meaning that our results are conservative.

\begin{table}[h]
\begin{center}
\input Tables/MNU_result.tex
 \caption{\label{tab:fit}
Upper limit (95\% confidence) on the neutrino mass (in \ev) in the \planck\
   Bayesian  framework and in the frequentist one based on
   Feldman--Cousins prescription.}
\end{center}
\end{table}
  

\section{Conclusion}

The use of Bayesian methodology in cosmology is partly motivated by the fact that one
observes a single realization of the Universe,  
while, in particle physics, one accumulates a number of events which
leads more naturally to using frequentist methods.
This argument is of a sociological rather than scientific nature, and nothing prevents
us from using one or the other methodology in these fields.

We demonstrated that a purely frequentist method is
tractable with the recent \planck-led high-precision cosmology data.
It required lowering the numerical noise of the Boltzmann solver
code and we have provided a set of precision parameters for the \class
software that, in
conjunction with a proper \minuit minimization strategy, allowed us to
perform the roughly $40$ parameter optimization efficiently.
We re-determined the maximum likelihood solution, obtaining
essentially consistent results but with a slightly better $\chi^2$ value.

We built profile likelihoods for each of the cosmological parameters
of the \lcdm\ model, using
the \cmb and \cmbao data sets, and obtained results very similar 
to those from the Bayesian methodology. This confirmed, in this model, that the \planck\
results do not depend on the choice of base parameters (implicit
priors) and are free of volume effects in the likelihood projection during the
marginalization process.

When including the neutrino mass as a free parameter,
the profile likelihood helped us to understand why the computed upper limit
increases when including the extra information from CMB lensing.
This is not due to the Bayesian methodology, but is
related to the physical boundary $\mnu>0$.
The profile likelihood analysis showed that neutrino mass limits
obtained without using the lensing information were pulled
down to unphysical negative values. Including the extra
CMB lensing information allowed us to obtain consistent frequentist results.

Using the Feldman--Cousins prescription, we obtained a 95\% confidence
upper limit of $\mnu \le 0.26\ev$ for the CMB+lensing+BAO combination,
again in excellent agreement with the Bayesian result.

\input acknowledgements

\appendix
\section{Note on CPU time}

It it sometimes stated that multi-dimensional minimization in
high-dimension space is inefficient (or intractable) while MCMC
methods scale linearly. Both statements need clarification.

Standard MCMC methods (e.g., Metropolis--Hastings or Gibbs sampling as in
\COSMOMC) are extremely CPU-intensive. They require the
lengthy computation of a multi-variate proposal before running a final
Markov chain, which by essence is sequential and therefore cannot scale on
multiple processors. In the \planck\ case about ${\cal O}(10^ 5)$
iterations (\ie, computations of the likelihood) were needed for this
final stage.

One \minuit minimization in our scheme is obtained in about ${\cal
  O}(10^ 4)$ iterations. It, however, requires a higher precision
tuning of the Boltzmann solver, which enhances the computation time of each likelihood by about
a factor two. In practice the minimum, in the $D=40$ case, is found in
about 10 hours, and is limited by the Boltzmann computation speed.
The profile likelihood approach requires many minimizations but these are
independent of one another. The problem now scales with the
number of computers, so that the total wall-clock time is still of the same order
of magnitude on a reasonable computer cluster.

\input references
\raggedright
\end{document}

%% file: macros.tex
\def\WMAP{\textit{WMAP}}

\newcommand{\WP}{WP}

\newcommand{\As}{A_{\rm s}}

\newcommand{\ns}{n_{\rm s}}
\newcommand{\lcdm}{$\Lambda$CDM}

\newcommand{\Alens}{A_{\rm L}}

\newcommand{\zre}{z_{\text{re}}}

\newcommand{\mnu}{\sum m_\nu}

\providecommand{\Planck}{\textit{Planck}}
\providecommand{\planck}{\Planck}

\providecommand{\text}[1]{\rm{#1}}

\providecommand{\muK}{\mu\rm{K}}

\providecommand{\Omb}{\Omega_{\mathrm{b}}}

\providecommand{\omb}{\omega_{\mathrm{b}}}
\providecommand{\omc}{\omega_{\mathrm{c}}}

\providecommand{\CAMB}{{\tt camb}}
\providecommand{\COSMOMC}{{\tt CosmoMC}}

\providecommand{\CLASS}{{\tt class}}
\providecommand{\MINUIT}{{\tt Minuit}}

\newcommand{\begm}{\begin{pmatrix}}
\newcommand{\enm}{\end{pmatrix}}

\newcommand\ba{\begin{eqnarray}}
\newcommand\ea{\end{eqnarray}}
\newcommand\bea{\begin{eqnarray}}
\newcommand\eea{\end{eqnarray}}

\newcommand\be{\begin{equation}}
\newcommand\ee{\end{equation}}

\def\pmb#1{\setbox0=\hbox{#1}%
    \kern-.025em\copy0\kern-\wd0
    \kern.05em\copy0\kern-\wd0
    \kern-.025em\raise.0433em\box0}
\def\ltsima{$\; \buildrel < \over \sim \;$}
\def\gtsima{$\; \buildrel > \over \sim \;$}
\def\simlt{\lower.5ex\hbox{\ltsima}}
\def\simgt{\lower.5ex\hbox{\gtsima}}

\def\p2Y{\;_2Y}
\def\m2Y{\;_{-2}Y}
\newcounter{parentequation}

\newcommand{\ev}{\ensuremath{\mathrm{\,e\kern -0.1em V}}\xspace}

\def\ie{{i.e.}\xspace}

\def\wrt{with respect to\xspace}

\def\to{\ensuremath{\rightarrow}\xspace}
\newcommand{\planckcoll}{\Planck\ Collaboration}
\newcommand{\wmap}{\WMAP\xspace}

\newcommand{\spt}{{SPT/ACT}\xspace}
\newcommand{\camb}{\CAMB\xspace}
\newcommand{\cosmomc}{\COSMOMC\xspace}
\newcommand{\minuit}{\MINUIT\xspace}
\newcommand{\class}{\CLASS\xspace}
\newcommand{\zreio}{\ensuremath{\zre}\xspace}

\newcommand{\dchi}{\ensuremath{\Delta \chi^2}\xspace}

\newcommand{\hz}{\ensuremath{H_0}\xspace}
\newcommand{\lnAs}{\ensuremath{\ln(10^{10} A_\mathrm{s})}\xspace}
\newcommand{\plancko}{\textsl{Planck-only}}
\newcommand{\cmb}{CMB\xspace}
\newcommand{\cmbao}{CMB+BAO\xspace}

\newcommand{\cl}{\ensuremath{C_\ell}\xspace}

\newcommand{\Al}{\ensuremath{\Alens\xspace}}

\def\deg{\degr}

\newcommand{\sect}[1]{Sect.~\ref{#1}\xspace}

%% file: Planck.tex
\def\setsymbol#1#2{\expandafter\def\csname #1\endcsname{#2}}
\def\getsymbol#1{\csname #1\endcsname}

\def\Planck{\textit{Planck}}

\newbox\tablebox    \newdimen\tablewidth
\def\leaderfil{\leaders\hbox to 5pt{\hss.\hss}\hfil}
\def\endPlancktable{\tablewidth=\columnwidth 
    $$\hss\copy\tablebox\hss$$
    \vskip-\lastskip\vskip -2pt}

\def\tablenote#1 #2\par{\begingroup \parindent=0.8em
    \abovedisplayshortskip=0pt\belowdisplayshortskip=0pt
    \noindent
    $$\hss\vbox{\hsize\tablewidth \hangindent=\parindent \hangafter=1 \noindent
    \hbox to \parindent{$^#1$\hss}\strut#2\strut\par}\hss$$
    \endgroup}
\def\doubleline{\vskip 3pt\hrule \vskip 1.5pt \hrule \vskip 5pt}

\def\L2{\ifmmode L_2\else $L_2$\fi}

\def\DeltaT{\ifmmode \Delta T\else $\Delta T$\fi}
\def\deltat{\ifmmode \Delta t\else $\Delta t$\fi}
\def\fknee{\ifmmode f_{\rm knee}\else $f_{\rm knee}$\fi}
\def\Fmax{\ifmmode F_{\rm max}\else $F_{\rm max}$\fi}
\def\solar{\ifmmode{\rm M}_{\mathord\odot}\else${\rm M}_{\mathord\odot}$\fi}
\def\Msolar{\ifmmode{\rm M}_{\mathord\odot}\else${\rm M}_{\mathord\odot}$\fi}
\def\Lsolar{\ifmmode{\rm L}_{\mathord\odot}\else${\rm L}_{\mathord\odot}$\fi}

\def\inv{\ifmmode^{-1}\else$^{-1}$\fi}
\def\mo{\ifmmode^{-1}\else$^{-1}$\fi}
\def\sup#1{\ifmmode ^{\rm #1}\else $^{\rm #1}$\fi}
\def\expo#1{\ifmmode \times 10^{#1}\else $\times 10^{#1}$\fi}
\def\,{\thinspace}
\def\lsim{\mathrel{\raise .4ex\hbox{\rlap{$<$}\lower 1.2ex\hbox{$\sim$}}}}
\def\gsim{\mathrel{\raise .4ex\hbox{\rlap{$>$}\lower 1.2ex\hbox{$\sim$}}}}

\def\simprop{\mathrel{\raise .4ex\hbox{\rlap{$\propto$}\lower 1.2ex\hbox{$\sim$}}}}
\def\deg{\ifmmode^\circ\else$^\circ$\fi}
\def\pdeg{\ifmmode $\setbox0=\hbox{$^{\circ}$}\rlap{\hskip.11\wd0 .}$^{\circ}
          \else \setbox0=\hbox{$^{\circ}$}\rlap{\hskip.11\wd0 .}$^{\circ}$\fi}
\def\arcs{\ifmmode {^{\scriptstyle\prime\prime}}
          \else $^{\scriptstyle\prime\prime}$\fi}
\def\arcm{\ifmmode {^{\scriptstyle\prime}}
          \else $^{\scriptstyle\prime}$\fi}
\newdimen\sa  \newdimen\sb
\def\parcs{\sa=.07em \sb=.03em
     \ifmmode \hbox{\rlap{.}}^{\scriptstyle\prime\kern -\sb\prime}\hbox{\kern -\sa}
     \else \rlap{.}$^{\scriptstyle\prime\kern -\sb\prime}$\kern -\sa\fi}
\def\parcm{\sa=.08em \sb=.03em
     \ifmmode \hbox{\rlap{.}\kern\sa}^{\scriptstyle\prime}\hbox{\kern-\sb}
     \else \rlap{.}\kern\sa$^{\scriptstyle\prime}$\kern-\sb\fi}
\def\ra[#1 #2 #3.#4]{#1\sup{h}#2\sup{m}#3\sup{s}\llap.#4}
\def\dec[#1 #2 #3.#4]{#1\deg#2\arcm#3\arcs\llap.#4}
\def\deco[#1 #2 #3]{#1\deg#2\arcm#3\arcs}
\def\rra[#1 #2]{#1\sup{h}#2\sup{m}}

\def\dots{\relax\ifmmode \ldots\else $\ldots$\fi}
\def\WHzsr{\ifmmode $W\,Hz\mo\,sr\mo$\else W\,Hz\mo\,sr\mo\fi}
\def\mHz{\ifmmode $\,mHz$\else \,mHz\fi}
\def\GHz{\ifmmode $\,GHz$\else \,GHz\fi}
\def\mKs{\ifmmode $\,mK\,s$^{1/2}\else \,mK\,s$^{1/2}$\fi}
\def\muKs{\ifmmode \,\mu$K\,s$^{1/2}\else \,$\mu$K\,s$^{1/2}$\fi}
\def\muKRJs{\ifmmode \,\mu$K$_{\rm RJ}$\,s$^{1/2}\else \,$\mu$K$_{\rm RJ}$\,s$^{1/2}$\fi}
\def\muKHz{\ifmmode \,\mu$K\,Hz$^{-1/2}\else \,$\mu$K\,Hz$^{-1/2}$\fi}
\def\MJysr{\ifmmode \,$MJy\,sr\mo$\else \,MJy\,sr\mo\fi}
\def\MJysrmK{\ifmmode \,$MJy\,sr\mo$\,mK$_{\rm CMB}\mo\else \,MJy\,sr\mo\,mK$_{\rm CMB}\mo$\fi}
\def\microns{\ifmmode \,\mu$m$\else \,$\mu$m\fi}

\def\muK{\ifmmode \,\mu$K$\else \,$\mu$\hbox{K}\fi}
\def\microK{\ifmmode \,\mu$K$\else \,$\mu$\hbox{K}\fi}
\def\muW{\ifmmode \,\mu$W$\else \,$\mu$\hbox{W}\fi}
\def\kms{\ifmmode $\,km\,s$^{-1}\else \,km\,s$^{-1}$\fi}
\def\kmsMpc{\ifmmode $\,\kms\,Mpc\mo$\else \,\kms\,Mpc\mo\fi}
\setsymbol{LFI:center:frequency:70GHz:units}{70.3\,GHz}
\setsymbol{LFI:center:frequency:44GHz:units}{44.1\,GHz}
\setsymbol{LFI:center:frequency:30GHz:units}{28.5\,GHz}

\setsymbol{LFI:center:frequency:70GHz}{70.3}
\setsymbol{LFI:center:frequency:44GHz}{44.1}
\setsymbol{LFI:center:frequency:30GHz}{28.5}

\setsymbol{LFI:center:frequency:LFI18:Rad:M:units}{71.7\GHz}
\setsymbol{LFI:center:frequency:LFI19:Rad:M:units}{67.5\GHz}
\setsymbol{LFI:center:frequency:LFI20:Rad:M:units}{69.2\GHz}
\setsymbol{LFI:center:frequency:LFI21:Rad:M:units}{70.4\GHz}
\setsymbol{LFI:center:frequency:LFI22:Rad:M:units}{71.5\GHz}
\setsymbol{LFI:center:frequency:LFI23:Rad:M:units}{70.8\GHz}
\setsymbol{LFI:center:frequency:LFI24:Rad:M:units}{44.4\GHz}
\setsymbol{LFI:center:frequency:LFI25:Rad:M:units}{44.0\GHz}
\setsymbol{LFI:center:frequency:LFI26:Rad:M:units}{43.9\GHz}
\setsymbol{LFI:center:frequency:LFI27:Rad:M:units}{28.3\GHz}
\setsymbol{LFI:center:frequency:LFI28:Rad:M:units}{28.8\GHz}
\setsymbol{LFI:center:frequency:LFI18:Rad:S:units}{70.1\GHz}
\setsymbol{LFI:center:frequency:LFI19:Rad:S:units}{69.6\GHz}
\setsymbol{LFI:center:frequency:LFI20:Rad:S:units}{69.5\GHz}
\setsymbol{LFI:center:frequency:LFI21:Rad:S:units}{69.5\GHz}
\setsymbol{LFI:center:frequency:LFI22:Rad:S:units}{72.8\GHz}
\setsymbol{LFI:center:frequency:LFI23:Rad:S:units}{71.3\GHz}
\setsymbol{LFI:center:frequency:LFI24:Rad:S:units}{44.1\GHz}
\setsymbol{LFI:center:frequency:LFI25:Rad:S:units}{44.1\GHz}
\setsymbol{LFI:center:frequency:LFI26:Rad:S:units}{44.1\GHz}
\setsymbol{LFI:center:frequency:LFI27:Rad:S:units}{28.5\GHz}
\setsymbol{LFI:center:frequency:LFI28:Rad:S:units}{28.2\GHz}

\setsymbol{LFI:center:frequency:LFI18:Rad:M}{71.7}
\setsymbol{LFI:center:frequency:LFI19:Rad:M}{67.5}
\setsymbol{LFI:center:frequency:LFI20:Rad:M}{69.2}
\setsymbol{LFI:center:frequency:LFI21:Rad:M}{70.4}
\setsymbol{LFI:center:frequency:LFI22:Rad:M}{71.5}
\setsymbol{LFI:center:frequency:LFI23:Rad:M}{70.8}
\setsymbol{LFI:center:frequency:LFI24:Rad:M}{44.4}
\setsymbol{LFI:center:frequency:LFI25:Rad:M}{44.0}
\setsymbol{LFI:center:frequency:LFI26:Rad:M}{43.9}
\setsymbol{LFI:center:frequency:LFI27:Rad:M}{28.3}
\setsymbol{LFI:center:frequency:LFI28:Rad:M}{28.8}
\setsymbol{LFI:center:frequency:LFI18:Rad:S}{70.1}
\setsymbol{LFI:center:frequency:LFI19:Rad:S}{69.6}
\setsymbol{LFI:center:frequency:LFI20:Rad:S}{69.5}
\setsymbol{LFI:center:frequency:LFI21:Rad:S}{69.5}
\setsymbol{LFI:center:frequency:LFI22:Rad:S}{72.8}
\setsymbol{LFI:center:frequency:LFI23:Rad:S}{71.3}
\setsymbol{LFI:center:frequency:LFI24:Rad:S}{44.1}
\setsymbol{LFI:center:frequency:LFI25:Rad:S}{44.1}
\setsymbol{LFI:center:frequency:LFI26:Rad:S}{44.1}
\setsymbol{LFI:center:frequency:LFI27:Rad:S}{28.5}
\setsymbol{LFI:center:frequency:LFI28:Rad:S}{28.2}

\setsymbol{LFI:white:noise:sensitivity:70GHz:units}{134.7\muKs}
\setsymbol{LFI:white:noise:sensitivity:44GHz:units}{164.7\muKs}
\setsymbol{LFI:white:noise:sensitivity:30GHz:units}{143.4\muKs}

\setsymbol{LFI:white:noise:sensitivity:70GHz}{134.7}
\setsymbol{LFI:white:noise:sensitivity:44GHz}{164.7}
\setsymbol{LFI:white:noise:sensitivity:30GHz}{143.4}

\setsymbol{LFI:white:noise:sensitivity:LFI18:Rad:M:units}{512.0\muKs}
\setsymbol{LFI:white:noise:sensitivity:LFI19:Rad:M:units}{581.4\muKs}
\setsymbol{LFI:white:noise:sensitivity:LFI20:Rad:M:units}{590.8\muKs}
\setsymbol{LFI:white:noise:sensitivity:LFI21:Rad:M:units}{455.2\muKs}
\setsymbol{LFI:white:noise:sensitivity:LFI22:Rad:M:units}{492.0\muKs}
\setsymbol{LFI:white:noise:sensitivity:LFI23:Rad:M:units}{507.7\muKs}
\setsymbol{LFI:white:noise:sensitivity:LFI24:Rad:M:units}{462.2\muKs}
\setsymbol{LFI:white:noise:sensitivity:LFI25:Rad:M:units}{413.6\muKs}
\setsymbol{LFI:white:noise:sensitivity:LFI26:Rad:M:units}{478.6\muKs}
\setsymbol{LFI:white:noise:sensitivity:LFI27:Rad:M:units}{277.7\muKs}
\setsymbol{LFI:white:noise:sensitivity:LFI28:Rad:M:units}{312.3\muKs}
\setsymbol{LFI:white:noise:sensitivity:LFI18:Rad:S:units}{465.7\muKs}
\setsymbol{LFI:white:noise:sensitivity:LFI19:Rad:S:units}{555.6\muKs}
\setsymbol{LFI:white:noise:sensitivity:LFI20:Rad:S:units}{623.2\muKs}
\setsymbol{LFI:white:noise:sensitivity:LFI21:Rad:S:units}{564.1\muKs}
\setsymbol{LFI:white:noise:sensitivity:LFI22:Rad:S:units}{534.4\muKs}
\setsymbol{LFI:white:noise:sensitivity:LFI23:Rad:S:units}{542.4\muKs}
\setsymbol{LFI:white:noise:sensitivity:LFI24:Rad:S:units}{399.2\muKs}
\setsymbol{LFI:white:noise:sensitivity:LFI25:Rad:S:units}{392.6\muKs}
\setsymbol{LFI:white:noise:sensitivity:LFI26:Rad:S:units}{418.6\muKs}
\setsymbol{LFI:white:noise:sensitivity:LFI27:Rad:S:units}{302.9\muKs}
\setsymbol{LFI:white:noise:sensitivity:LFI28:Rad:S:units}{285.3\muKs}

\setsymbol{LFI:white:noise:sensitivity:LFI18:Rad:M}{512.0}
\setsymbol{LFI:white:noise:sensitivity:LFI19:Rad:M}{581.4}
\setsymbol{LFI:white:noise:sensitivity:LFI20:Rad:M}{590.8}
\setsymbol{LFI:white:noise:sensitivity:LFI21:Rad:M}{455.2}
\setsymbol{LFI:white:noise:sensitivity:LFI22:Rad:M}{492.0}
\setsymbol{LFI:white:noise:sensitivity:LFI23:Rad:M}{507.7}
\setsymbol{LFI:white:noise:sensitivity:LFI24:Rad:M}{462.2}
\setsymbol{LFI:white:noise:sensitivity:LFI25:Rad:M}{413.6}
\setsymbol{LFI:white:noise:sensitivity:LFI26:Rad:M}{478.6}
\setsymbol{LFI:white:noise:sensitivity:LFI27:Rad:M}{277.7}
\setsymbol{LFI:white:noise:sensitivity:LFI28:Rad:M}{312.3}
\setsymbol{LFI:white:noise:sensitivity:LFI18:Rad:S}{465.7}
\setsymbol{LFI:white:noise:sensitivity:LFI19:Rad:S}{555.6}
\setsymbol{LFI:white:noise:sensitivity:LFI20:Rad:S}{623.2}
\setsymbol{LFI:white:noise:sensitivity:LFI21:Rad:S}{564.1}
\setsymbol{LFI:white:noise:sensitivity:LFI22:Rad:S}{534.4}
\setsymbol{LFI:white:noise:sensitivity:LFI23:Rad:S}{542.4}
\setsymbol{LFI:white:noise:sensitivity:LFI24:Rad:S}{399.2}
\setsymbol{LFI:white:noise:sensitivity:LFI25:Rad:S}{392.6}
\setsymbol{LFI:white:noise:sensitivity:LFI26:Rad:S}{418.6}
\setsymbol{LFI:white:noise:sensitivity:LFI27:Rad:S}{302.9}
\setsymbol{LFI:white:noise:sensitivity:LFI28:Rad:S}{285.3}

\setsymbol{LFI:knee:frequency:70GHz:units}{29.5\mHz}
\setsymbol{LFI:knee:frequency:44GHz:units}{56.2\mHz}
\setsymbol{LFI:knee:frequency:30GHz:units}{113.7\mHz}

\setsymbol{LFI:knee:frequency:70GHz}{29.5}
\setsymbol{LFI:knee:frequency:44GHz}{56.2}
\setsymbol{LFI:knee:frequency:30GHz}{113.7}

\setsymbol{LFI:knee:frequency:LFI18:Rad:M:units}{16.3\mHz}
\setsymbol{LFI:knee:frequency:LFI19:Rad:M:units}{15.1\mHz}
\setsymbol{LFI:knee:frequency:LFI20:Rad:M:units}{18.7\mHz}
\setsymbol{LFI:knee:frequency:LFI21:Rad:M:units}{37.2\mHz}
\setsymbol{LFI:knee:frequency:LFI22:Rad:M:units}{12.7\mHz}
\setsymbol{LFI:knee:frequency:LFI23:Rad:M:units}{34.6\mHz}
\setsymbol{LFI:knee:frequency:LFI24:Rad:M:units}{46.2\mHz}
\setsymbol{LFI:knee:frequency:LFI25:Rad:M:units}{24.9\mHz}
\setsymbol{LFI:knee:frequency:LFI26:Rad:M:units}{67.6\mHz}
\setsymbol{LFI:knee:frequency:LFI27:Rad:M:units}{187.4\mHz}
\setsymbol{LFI:knee:frequency:LFI28:Rad:M:units}{122.2\mHz}
\setsymbol{LFI:knee:frequency:LFI18:Rad:S:units}{17.7\mHz}
\setsymbol{LFI:knee:frequency:LFI19:Rad:S:units}{22.0\mHz}
\setsymbol{LFI:knee:frequency:LFI20:Rad:S:units}{8.7\mHz}
\setsymbol{LFI:knee:frequency:LFI21:Rad:S:units}{25.9\mHz}
\setsymbol{LFI:knee:frequency:LFI22:Rad:S:units}{15.8\mHz}
\setsymbol{LFI:knee:frequency:LFI23:Rad:S:units}{129.8\mHz}
\setsymbol{LFI:knee:frequency:LFI24:Rad:S:units}{100.9\mHz}
\setsymbol{LFI:knee:frequency:LFI25:Rad:S:units}{38.9\mHz}
\setsymbol{LFI:knee:frequency:LFI26:Rad:S:units}{58.9\mHz}
\setsymbol{LFI:knee:frequency:LFI27:Rad:S:units}{104.4\mHz}
\setsymbol{LFI:knee:frequency:LFI28:Rad:S:units}{40.7\mHz}

\setsymbol{LFI:knee:frequency:LFI18:Rad:M}{16.3}
\setsymbol{LFI:knee:frequency:LFI19:Rad:M}{15.1}
\setsymbol{LFI:knee:frequency:LFI20:Rad:M}{18.7}
\setsymbol{LFI:knee:frequency:LFI21:Rad:M}{37.2}
\setsymbol{LFI:knee:frequency:LFI22:Rad:M}{12.7}
\setsymbol{LFI:knee:frequency:LFI23:Rad:M}{34.6}
\setsymbol{LFI:knee:frequency:LFI24:Rad:M}{46.2}
\setsymbol{LFI:knee:frequency:LFI25:Rad:M}{24.9}
\setsymbol{LFI:knee:frequency:LFI26:Rad:M}{67.6}
\setsymbol{LFI:knee:frequency:LFI27:Rad:M}{187.4}
\setsymbol{LFI:knee:frequency:LFI28:Rad:M}{122.2}
\setsymbol{LFI:knee:frequency:LFI18:Rad:S}{17.7}
\setsymbol{LFI:knee:frequency:LFI19:Rad:S}{22.0}
\setsymbol{LFI:knee:frequency:LFI20:Rad:S}{8.7}
\setsymbol{LFI:knee:frequency:LFI21:Rad:S}{25.9}
\setsymbol{LFI:knee:frequency:LFI22:Rad:S}{15.8}
\setsymbol{LFI:knee:frequency:LFI23:Rad:S}{129.8}
\setsymbol{LFI:knee:frequency:LFI24:Rad:S}{100.9}
\setsymbol{LFI:knee:frequency:LFI25:Rad:S}{38.9}
\setsymbol{LFI:knee:frequency:LFI26:Rad:S}{58.9}
\setsymbol{LFI:knee:frequency:LFI27:Rad:S}{104.4}
\setsymbol{LFI:knee:frequency:LFI28:Rad:S}{40.7}

\setsymbol{LFI:slope:70GHz:units}{$-1.03$\mHz}
\setsymbol{LFI:slope:44GHz:units}{$-0.89$\mHz}
\setsymbol{LFI:slope:30GHz:units}{$-0.87$\mHz}

\setsymbol{LFI:slope:70GHz}{$-1.03$}
\setsymbol{LFI:slope:44GHz}{$-0.89$}
\setsymbol{LFI:slope:30GHz}{$-0.87$}

\setsymbol{LFI:slope:LFI18:Rad:M:units}{$-1.04$\mHz}
\setsymbol{LFI:slope:LFI19:Rad:M:units}{$-1.09$\mHz}
\setsymbol{LFI:slope:LFI20:Rad:M:units}{$-0.69$\mHz}
\setsymbol{LFI:slope:LFI21:Rad:M:units}{$-1.56$\mHz}
\setsymbol{LFI:slope:LFI22:Rad:M:units}{$-1.01$\mHz}
\setsymbol{LFI:slope:LFI23:Rad:M:units}{$-0.96$\mHz}
\setsymbol{LFI:slope:LFI24:Rad:M:units}{$-0.83$\mHz}
\setsymbol{LFI:slope:LFI25:Rad:M:units}{$-0.91$\mHz}
\setsymbol{LFI:slope:LFI26:Rad:M:units}{$-0.95$\mHz}
\setsymbol{LFI:slope:LFI27:Rad:M:units}{$-0.87$\mHz}
\setsymbol{LFI:slope:LFI28:Rad:M:units}{$-0.88$\mHz}
\setsymbol{LFI:slope:LFI18:Rad:S:units}{$-1.15$\mHz}
\setsymbol{LFI:slope:LFI19:Rad:S:units}{$-1.00$\mHz}
\setsymbol{LFI:slope:LFI20:Rad:S:units}{$-0.95$\mHz}
\setsymbol{LFI:slope:LFI21:Rad:S:units}{$-0.92$\mHz}
\setsymbol{LFI:slope:LFI22:Rad:S:units}{$-1.01$\mHz}
\setsymbol{LFI:slope:LFI23:Rad:S:units}{$-0.95$\mHz}
\setsymbol{LFI:slope:LFI24:Rad:S:units}{$-0.73$\mHz}
\setsymbol{LFI:slope:LFI25:Rad:S:units}{$-1.16$\mHz}
\setsymbol{LFI:slope:LFI26:Rad:S:units}{$-0.79$\mHz}
\setsymbol{LFI:slope:LFI27:Rad:S:units}{$-0.82$\mHz}
\setsymbol{LFI:slope:LFI28:Rad:S:units}{$-0.91$\mHz}

\setsymbol{LFI:slope:LFI18:Rad:M}{$-1.04$}
\setsymbol{LFI:slope:LFI19:Rad:M}{$-1.09$}
\setsymbol{LFI:slope:LFI20:Rad:M}{$-0.69$}
\setsymbol{LFI:slope:LFI21:Rad:M}{$-1.56$}
\setsymbol{LFI:slope:LFI22:Rad:M}{$-1.01$}
\setsymbol{LFI:slope:LFI23:Rad:M}{$-0.96$}
\setsymbol{LFI:slope:LFI24:Rad:M}{$-0.83$}
\setsymbol{LFI:slope:LFI25:Rad:M}{$-0.91$}
\setsymbol{LFI:slope:LFI26:Rad:M}{$-0.95$}
\setsymbol{LFI:slope:LFI27:Rad:M}{$-0.87$}
\setsymbol{LFI:slope:LFI28:Rad:M}{$-0.88$}
\setsymbol{LFI:slope:LFI18:Rad:S}{$-1.15$}
\setsymbol{LFI:slope:LFI19:Rad:S}{$-1.00$}
\setsymbol{LFI:slope:LFI20:Rad:S}{$-0.95$}
\setsymbol{LFI:slope:LFI21:Rad:S}{$-0.92$}
\setsymbol{LFI:slope:LFI22:Rad:S}{$-1.01$}
\setsymbol{LFI:slope:LFI23:Rad:S}{$-0.95$}
\setsymbol{LFI:slope:LFI24:Rad:S}{$-0.73$}
\setsymbol{LFI:slope:LFI25:Rad:S}{$-1.16$}
\setsymbol{LFI:slope:LFI26:Rad:S}{$-0.79$}
\setsymbol{LFI:slope:LFI27:Rad:S}{$-0.82$}
\setsymbol{LFI:slope:LFI28:Rad:S}{$-0.91$}

\setsymbol{LFI:FWHM:70GHz:units}{13\parcm01}
\setsymbol{LFI:FWHM:44GHz:units}{27\parcm92}
\setsymbol{LFI:FWHM:30GHz:units}{32\parcm65}

\setsymbol{LFI:FWHM:70GHz}{13.01}
\setsymbol{LFI:FWHM:44GHz}{27.92}
\setsymbol{LFI:FWHM:30GHz}{32.65}

\setsymbol{LFI:FWHM:LFI18:units}{13\parcm39}
\setsymbol{LFI:FWHM:LFI19:units}{13\parcm01}
\setsymbol{LFI:FWHM:LFI20:units}{12\parcm75}
\setsymbol{LFI:FWHM:LFI21:units}{12\parcm74}
\setsymbol{LFI:FWHM:LFI22:units}{12\parcm87}
\setsymbol{LFI:FWHM:LFI23:units}{13\parcm27}
\setsymbol{LFI:FWHM:LFI24:units}{22\parcm98}
\setsymbol{LFI:FWHM:LFI25:units}{30\parcm46}
\setsymbol{LFI:FWHM:LFI26:units}{30\parcm31}
\setsymbol{LFI:FWHM:LFI27:units}{32\parcm65}
\setsymbol{LFI:FWHM:LFI28:units}{32\parcm66}

\setsymbol{LFI:FWHM:LFI18}{13.39}
\setsymbol{LFI:FWHM:LFI19}{13.01}
\setsymbol{LFI:FWHM:LFI20}{12.75}
\setsymbol{LFI:FWHM:LFI21}{12.74}
\setsymbol{LFI:FWHM:LFI22}{12.87}
\setsymbol{LFI:FWHM:LFI23}{13.27}
\setsymbol{LFI:FWHM:LFI24}{22.98}
\setsymbol{LFI:FWHM:LFI25}{30.46}
\setsymbol{LFI:FWHM:LFI26}{30.31}
\setsymbol{LFI:FWHM:LFI27}{32.65}
\setsymbol{LFI:FWHM:LFI28}{32.66}

\setsymbol{LFI:FWHM:uncertainty:LFI18:units}{0.170\arcm}
\setsymbol{LFI:FWHM:uncertainty:LFI19:units}{0.174\arcm}
\setsymbol{LFI:FWHM:uncertainty:LFI20:units}{0.170\arcm}
\setsymbol{LFI:FWHM:uncertainty:LFI21:units}{0.156\arcm}
\setsymbol{LFI:FWHM:uncertainty:LFI22:units}{0.164\arcm}
\setsymbol{LFI:FWHM:uncertainty:LFI23:units}{0.171\arcm}
\setsymbol{LFI:FWHM:uncertainty:LFI24:units}{0.652\arcm}
\setsymbol{LFI:FWHM:uncertainty:LFI25:units}{1.075\arcm}
\setsymbol{LFI:FWHM:uncertainty:LFI26:units}{1.131\arcm}
\setsymbol{LFI:FWHM:uncertainty:LFI27:units}{1.266\arcm}
\setsymbol{LFI:FWHM:uncertainty:LFI28:units}{1.287\arcm}

\setsymbol{LFI:FWHM:uncertainty:LFI18}{0.170}
\setsymbol{LFI:FWHM:uncertainty:LFI19}{0.174}
\setsymbol{LFI:FWHM:uncertainty:LFI20}{0.170}
\setsymbol{LFI:FWHM:uncertainty:LFI21}{0.156}
\setsymbol{LFI:FWHM:uncertainty:LFI22}{0.164}
\setsymbol{LFI:FWHM:uncertainty:LFI23}{0.171}
\setsymbol{LFI:FWHM:uncertainty:LFI24}{0.652}
\setsymbol{LFI:FWHM:uncertainty:LFI25}{1.075}
\setsymbol{LFI:FWHM:uncertainty:LFI26}{1.131}
\setsymbol{LFI:FWHM:uncertainty:LFI27}{1.266}
\setsymbol{LFI:FWHM:uncertainty:LFI28}{1.287}

\setsymbol{HFI:center:frequency:100GHz:units}{100\,GHz}
\setsymbol{HFI:center:frequency:143GHz:units}{143\,GHz}
\setsymbol{HFI:center:frequency:217GHz:units}{217\,GHz}
\setsymbol{HFI:center:frequency:353GHz:units}{353\,GHz}
\setsymbol{HFI:center:frequency:545GHz:units}{545\,GHz}
\setsymbol{HFI:center:frequency:857GHz:units}{857\,GHz}

\setsymbol{HFI:center:frequency:100GHz}{100}
\setsymbol{HFI:center:frequency:143GHz}{143}
\setsymbol{HFI:center:frequency:217GHz}{217}
\setsymbol{HFI:center:frequency:353GHz}{353}
\setsymbol{HFI:center:frequency:545GHz}{545}
\setsymbol{HFI:center:frequency:857GHz}{857}

\setsymbol{HFI:Ndetectors:100GHz}{8}
\setsymbol{HFI:Ndetectors:143GHz}{11}
\setsymbol{HFI:Ndetectors:217GHz}{12}
\setsymbol{HFI:Ndetectors:353GHz}{12}
\setsymbol{HFI:Ndetectors:545GHz}{3}
\setsymbol{HFI:Ndetectors:857GHz}{4}

\setsymbol{HFI:FWHM:Maps:100GHz:units}{9\parcm88}
\setsymbol{HFI:FWHM:Maps:143GHz:units}{7\parcm18}
\setsymbol{HFI:FWHM:Maps:217GHz:units}{4\parcm87}
\setsymbol{HFI:FWHM:Maps:353GHz:units}{4\parcm65}
\setsymbol{HFI:FWHM:Maps:545GHz:units}{4\parcm72}
\setsymbol{HFI:FWHM:Maps:857GHz:units}{4\parcm39}
\setsymbol{HFI:FWHM:Maps:100GHz}{9.88}
\setsymbol{HFI:FWHM:Maps:143GHz}{7.18}
\setsymbol{HFI:FWHM:Maps:217GHz}{4.87}
\setsymbol{HFI:FWHM:Maps:353GHz}{4.65}
\setsymbol{HFI:FWHM:Maps:545GHz}{4.72}
\setsymbol{HFI:FWHM:Maps:857GHz}{4.39}

\setsymbol{HFI:beam:ellipticity:Maps:100GHz}{1.15}
\setsymbol{HFI:beam:ellipticity:Maps:143GHz}{1.01}
\setsymbol{HFI:beam:ellipticity:Maps:217GHz}{1.06}
\setsymbol{HFI:beam:ellipticity:Maps:353GHz}{1.05}
\setsymbol{HFI:beam:ellipticity:Maps:545GHz}{1.14}
\setsymbol{HFI:beam:ellipticity:Maps:857GHz}{1.19}

\setsymbol{HFI:FWHM:Mars:100GHz:units}{9\parcm37}
\setsymbol{HFI:FWHM:Mars:143GHz:units}{7\parcm04}
\setsymbol{HFI:FWHM:Mars:217GHz:units}{4\parcm68}
\setsymbol{HFI:FWHM:Mars:353GHz:units}{4\parcm43}
\setsymbol{HFI:FWHM:Mars:545GHz:units}{3\parcm80}
\setsymbol{HFI:FWHM:Mars:857GHz:units}{3\parcm67}

\setsymbol{HFI:FWHM:Mars:100GHz}{9.37}
\setsymbol{HFI:FWHM:Mars:143GHz}{7.04}
\setsymbol{HFI:FWHM:Mars:217GHz}{4.68}
\setsymbol{HFI:FWHM:Mars:353GHz}{4.43}
\setsymbol{HFI:FWHM:Mars:545GHz}{3.80}
\setsymbol{HFI:FWHM:Mars:857GHz}{3.67}

\setsymbol{HFI:beam:ellipticity:Mars:100GHz}{1.18}
\setsymbol{HFI:beam:ellipticity:Mars:143GHz}{1.03}
\setsymbol{HFI:beam:ellipticity:Mars:217GHz}{1.14}
\setsymbol{HFI:beam:ellipticity:Mars:353GHz}{1.09}
\setsymbol{HFI:beam:ellipticity:Mars:545GHz}{1.25}
\setsymbol{HFI:beam:ellipticity:Mars:857GHz}{1.03}

\setsymbol{HFI:CMB:relative:calibration:100GHz}{$\lsim 1\%$}
\setsymbol{HFI:CMB:relative:calibration:143GHz}{$\lsim 1\%$}
\setsymbol{HFI:CMB:relative:calibration:217GHz}{$\lsim 1\%$}
\setsymbol{HFI:CMB:relative:calibration:353GHz}{$\lsim 1\%$}
\setsymbol{HFI:CMB:relative:calibration:545GHz}{}
\setsymbol{HFI:CMB:relative:calibration:857GHz}{}

\setsymbol{HFI:CMB:absolute:calibration:100GHz}{$\lsim 2\%$}
\setsymbol{HFI:CMB:absolute:calibration:143GHz}{$\lsim 2\%$}
\setsymbol{HFI:CMB:absolute:calibration:217GHz}{$\lsim 2\%$}
\setsymbol{HFI:CMB:absolute:calibration:353GHz}{$\lsim 2\%$}
\setsymbol{HFI:CMB:absolute:calibration:545GHz}{}
\setsymbol{HFI:CMB:absolute:calibration:857GHz}{}

\setsymbol{HFI:FIRAS:gain:calibration:accuracy:statistical:100GHz}{}
\setsymbol{HFI:FIRAS:gain:calibration:accuracy:statistical:143GHz}{}
\setsymbol{HFI:FIRAS:gain:calibration:accuracy:statistical:217GHz}{}
\setsymbol{HFI:FIRAS:gain:calibration:accuracy:statistical:353GHz}{2.5\%}
\setsymbol{HFI:FIRAS:gain:calibration:accuracy:statistical:545GHz}{1\%}
\setsymbol{HFI:FIRAS:gain:calibration:accuracy:statistical:857GHz}{0.5\%}

\setsymbol{HFI:FIRAS:gain:calibration:accuracy:systematic:100GHz}{}
\setsymbol{HFI:FIRAS:gain:calibration:accuracy:systematic:143GHz}{}
\setsymbol{HFI:FIRAS:gain:calibration:accuracy:systematic:217GHz}{}
\setsymbol{HFI:FIRAS:gain:calibration:accuracy:systematic:353GHz}{}
\setsymbol{HFI:FIRAS:gain:calibration:accuracy:systematic:545GHz}{7\%}
\setsymbol{HFI:FIRAS:gain:calibration:accuracy:systematic:857GHz}{7\%}

\setsymbol{HFI:FIRAS:zero:point:accuracy:100GHz:units}{0.8\MJysr}
\setsymbol{HFI:FIRAS:zero:point:accuracy:143GHz:units}{}
\setsymbol{HFI:FIRAS:zero:point:accuracy:217GHz:units}{}
\setsymbol{HFI:FIRAS:zero:point:accuracy:353GHz:units}{1.4\MJysr}
\setsymbol{HFI:FIRAS:zero:point:accuracy:545GHz:units}{2.2\MJysr}
\setsymbol{HFI:FIRAS:zero:point:accuracy:857GHz:units}{1.7\MJysr}

\setsymbol{HFI:FIRAS:zero:point:accuracy:100GHz}{0.8}
\setsymbol{HFI:FIRAS:zero:point:accuracy:143GHz}{}
\setsymbol{HFI:FIRAS:zero:point:accuracy:217GHz}{}
\setsymbol{HFI:FIRAS:zero:point:accuracy:353GHz}{1.4}
\setsymbol{HFI:FIRAS:zero:point:accuracy:545GHz}{2.2}
\setsymbol{HFI:FIRAS:zero:point:accuracy:857GHz}{1.7}

\setsymbol{HFI:unit:conversion:100GHz:units}{0.2415\MJysrmK}
\setsymbol{HFI:unit:conversion:143GHz:units}{0.3694\MJysrmK}
\setsymbol{HFI:unit:conversion:217GHz:units}{0.4811\MJysrmK}
\setsymbol{HFI:unit:conversion:353GHz:units}{0.2883\MJysrmK}
\setsymbol{HFI:unit:conversion:545GHz:units}{0.05826\MJysrmK}
\setsymbol{HFI:unit:conversion:857GHz:units}{0.002238\MJysrmK}

\setsymbol{HFI:unit:conversion:100GHz}{0.2415}
\setsymbol{HFI:unit:conversion:143GHz}{0.3694}
\setsymbol{HFI:unit:conversion:217GHz}{0.4811}
\setsymbol{HFI:unit:conversion:353GHz}{0.2883}
\setsymbol{HFI:unit:conversion:545GHz}{0.05826}
\setsymbol{HFI:unit:conversion:857GHz}{0.002238}

\setsymbol{HFI:colour:correction:alpha=-2:V1.01:100GHz}{0.9893}
\setsymbol{HFI:colour:correction:alpha=-2:V1.01:143GHz}{0.9759}
\setsymbol{HFI:colour:correction:alpha=-2:V1.01:217GHz}{1.0007}
\setsymbol{HFI:colour:correction:alpha=-2:V1.01:353GHz}{1.0028}
\setsymbol{HFI:colour:correction:alpha=-2:V1.01:545GHz}{1.0019}
\setsymbol{HFI:colour:correction:alpha=-2:V1.01:857GHz}{0.9889}

\setsymbol{HFI:colour:correction:alpha=0:V1.01:100GHz}{1.0008}
\setsymbol{HFI:colour:correction:alpha=0:V1.01:143GHz}{1.0148}
\setsymbol{HFI:colour:correction:alpha=0:V1.01:217GHz}{0.9909}
\setsymbol{HFI:colour:correction:alpha=0:V1.01:353GHz}{0.9888}
\setsymbol{HFI:colour:correction:alpha=0:V1.01:545GHz}{0.9878}
\setsymbol{HFI:colour:correction:alpha=0:V1.01:857GHz}{1.0014}

%% file: title.tex
\title{\Planck\ intermediate results. XVI. \\Profile likelihoods for cosmological parameters}
\titlerunning{Profile likelihoods for cosmological parameters}
\authorrunning{Planck Collaboration}
\author{The Planck team}
\institute{L2 \and Earth}
\date{Received 1 January 2013/Accepted 1 January 2013}

%% file: P11a_Parameters_Frequentist_Estimation_authors_and_institutes.tex
\author{\small
Planck Collaboration:
P.~A.~R.~Ade\inst{79}
\and
N.~Aghanim\inst{56}
\and
M.~Arnaud\inst{68}
\and
M.~Ashdown\inst{65, 6}
\and
J.~Aumont\inst{56}
\and
C.~Baccigalupi\inst{77}
\and
A.~J.~Banday\inst{81, 10}
\and
R.~B.~Barreiro\inst{62}
\and
J.~G.~Bartlett\inst{1, 63}
\and
E.~Battaner\inst{82}
% ...
}
\institute{\small
APC, AstroParticule et Cosmologie, Universit\'{e} Paris Diderot, CNRS/IN2P3, CEA/lrfu, Observatoire de Paris, Sorbonne Paris Cit\'{e}, 10, rue Alice Domon et L\'{e}onie Duquet, 75205 Paris Cedex 13, France\\
\and
Aalto University Mets\"{a}hovi Radio Observatory and Dept of Radio Science and Engineering, P.O. Box 13000, FI-00076 AALTO, Finland\\
\and
African Institute for Mathematical Sciences, 6-8 Melrose Road, Muizenberg, Cape Town, South Africa\\
\and
Agenzia Spaziale Italiana Science Data Center, Via del Politecnico snc, 00133, Roma, Italy\\
\and
Agenzia Spaziale Italiana, Viale Liegi 26, Roma, Italy\\
\and
Astrophysics Group, Cavendish Laboratory, University of Cambridge, J J Thomson Avenue, Cambridge CB3 0HE, U.K.\\
\and
Astrophysics \& Cosmology Research Unit, School of Mathematics, Statistics \& Computer Science, University of KwaZulu-Natal, Westville Campus, Private Bag X54001, Durban 4000, South Africa\\
\and
Atacama Large Millimeter/submillimeter Array, ALMA Santiago Central Offices, Alonso de Cordova 3107, Vitacura, Casilla 763 0355, Santiago, Chile\\
\and
CITA, University of Toronto, 60 St. George St., Toronto, ON M5S 3H8, Canada\\
\and
CNRS, IRAP, 9 Av. colonel Roche, BP 44346, F-31028 Toulouse cedex 4, France\\
\and
California Institute of Technology, Pasadena, California, U.S.A.\\
\and
Centro de Estudios de F\'{i}sica del Cosmos de Arag\'{o}n (CEFCA), Plaza San Juan, 1, planta 2, E-44001, Teruel, Spain\\
\and
Computational Cosmology Center, Lawrence Berkeley National Laboratory, Berkeley, California, U.S.A.\\
\and
Consejo Superior de Investigaciones Cient\'{\i}ficas (CSIC), Madrid, Spain\\
\and
DSM/Irfu/SPP, CEA-Saclay, F-91191 Gif-sur-Yvette Cedex, France\\
\and
DTU Space, National Space Institute, Technical University of Denmark, Elektrovej 327, DK-2800 Kgs. Lyngby, Denmark\\
\and
D\'{e}partement de Physique Th\'{e}orique, Universit\'{e} de Gen\`{e}ve, 24, Quai E. Ansermet,1211 Gen\`{e}ve 4, Switzerland\\
\and
Departamento de F\'{\i}sica, Universidad de Oviedo, Avda. Calvo Sotelo s/n, Oviedo, Spain\\
\and
Department of Astrophysics/IMAPP, Radboud University Nijmegen, P.O. Box 9010, 6500 GL Nijmegen, The Netherlands\\
\and
Department of Electrical Engineering and Computer Sciences, University of California, Berkeley, California, U.S.A.\\
\and
Department of Physics and Astronomy, Dana and David Dornsife College of Letter, Arts and Sciences, University of Southern California, Los Angeles, CA 90089, U.S.A.\\
\and
Department of Physics and Astronomy, University College London, London WC1E 6BT, U.K.\\
\and
Department of Physics and Astronomy, University of Sussex, Brighton BN1 9QH, U.K.\\
\and
Department of Physics, Florida State University, Keen Physics Building, 77 Chieftan Way, Tallahassee, Florida, U.S.A.\\
\and
Department of Physics, Gustaf H\"{a}llstr\"{o}min katu 2a, University of Helsinki, Helsinki, Finland\\
\and
Department of Physics, Princeton University, Princeton, New Jersey, U.S.A.\\
\and
Department of Physics, University of California, Berkeley, California, U.S.A.\\
\and
Department of Physics, University of California, One Shields Avenue, Davis, California, U.S.A.\\
\and
Department of Physics, University of California, Santa Barbara, California, U.S.A.\\
\and
Department of Physics, University of Illinois at Urbana-Champaign, 1110 West Green Street, Urbana, Illinois, U.S.A.\\
\and
Dipartimento di Fisica e Astronomia G. Galilei, Universit\`{a} degli Studi di Padova, via Marzolo 8, 35131 Padova, Italy\\
\and
Dipartimento di Fisica e Scienze della Terra, Universit\`{a} di Ferrara, Via Saragat 1, 44122 Ferrara, Italy\\
\and
Dipartimento di Fisica, Universit\`{a} La Sapienza, P. le A. Moro 2, Roma, Italy\\
\and
Dipartimento di Fisica, Universit\`{a} degli Studi di Milano, Via Celoria, 16, Milano, Italy\\
\and
Dipartimento di Fisica, Universit\`{a} degli Studi di Trieste, via A. Valerio 2, Trieste, Italy\\
\and
Dipartimento di Fisica, Universit\`{a} di Roma Tor Vergata, Via della Ricerca Scientifica, 1, Roma, Italy\\
\and
Discovery Center, Niels Bohr Institute, Blegdamsvej 17, Copenhagen, Denmark\\
\and
Dpto. Astrof\'{i}sica, Universidad de La Laguna (ULL), E-38206 La Laguna, Tenerife, Spain\\
\and
European Southern Observatory, ESO Vitacura, Alonso de Cordova 3107, Vitacura, Casilla 19001, Santiago, Chile\\
\and
European Space Agency, ESAC, Planck Science Office, Camino bajo del Castillo, s/n, Urbanizaci\'{o}n Villafranca del Castillo, Villanueva de la Ca\~{n}ada, Madrid, Spain\\
\and
European Space Agency, ESTEC, Keplerlaan 1, 2201 AZ Noordwijk, The Netherlands\\
\and
Helsinki Institute of Physics, Gustaf H\"{a}llstr\"{o}min katu 2, University of Helsinki, Helsinki, Finland\\
\and
INAF - Osservatorio Astrofisico di Catania, Via S. Sofia 78, Catania, Italy\\
\and
INAF - Osservatorio Astronomico di Padova, Vicolo dell'Osservatorio 5, Padova, Italy\\
\and
INAF - Osservatorio Astronomico di Roma, via di Frascati 33, Monte Porzio Catone, Italy\\
\and
INAF - Osservatorio Astronomico di Trieste, Via G.B. Tiepolo 11, Trieste, Italy\\
\and
INAF Istituto di Radioastronomia, Via P. Gobetti 101, 40129 Bologna, Italy\\
\and
INAF/IASF Bologna, Via Gobetti 101, Bologna, Italy\\
\and
INAF/IASF Milano, Via E. Bassini 15, Milano, Italy\\
\and
INFN, Sezione di Bologna, Via Irnerio 46, I-40126, Bologna, Italy\\
\and
INFN, Sezione di Roma 1, Universit\`{a} di Roma Sapienza, Piazzale Aldo Moro 2, 00185, Roma, Italy\\
\and
IUCAA, Post Bag 4, Ganeshkhind, Pune University Campus, Pune 411 007, India\\
\and
Imperial College London, Astrophysics group, Blackett Laboratory, Prince Consort Road, London, SW7 2AZ, U.K.\\
\and
Infrared Processing and Analysis Center, California Institute of Technology, Pasadena, CA 91125, U.S.A.\\
\and
Institut Universitaire de France, 103, bd Saint-Michel, 75005, Paris, France\\
\and
Institut d'Astrophysique Spatiale, CNRS (UMR8617) Universit\'{e} Paris-Sud 11, B\^{a}timent 121, Orsay, France\\
\and
Institut d'Astrophysique de Paris, CNRS (UMR7095), 98 bis Boulevard Arago, F-75014, Paris, France\\
\and
Institute for Space Sciences, Bucharest-Magurale, Romania\\
\and
Institute of Astronomy, University of Cambridge, Madingley Road, Cambridge CB3 0HA, U.K.\\
\and
Institute of Theoretical Astrophysics, University of Oslo, Blindern, Oslo, Norway\\
\and
Instituto de Astrof\'{\i}sica de Canarias, C/V\'{\i}a L\'{a}ctea s/n, La Laguna, Tenerife, Spain\\
\and
Instituto de F\'{\i}sica de Cantabria (CSIC-Universidad de Cantabria), Avda. de los Castros s/n, Santander, Spain\\
\and
Jet Propulsion Laboratory, California Institute of Technology, 4800 Oak Grove Drive, Pasadena, California, U.S.A.\\
\and
Jodrell Bank Centre for Astrophysics, Alan Turing Building, School of Physics and Astronomy, The University of Manchester, Oxford Road, Manchester, M13 9PL, U.K.\\
\and
Kavli Institute for Cosmology Cambridge, Madingley Road, Cambridge, CB3 0HA, U.K.\\
\and
LAL, Universit\'{e} Paris-Sud, CNRS/IN2P3, Orsay, France\\
\and
LERMA, CNRS, Observatoire de Paris, 61 Avenue de l'Observatoire, Paris, France\\
\and
Laboratoire AIM, IRFU/Service d'Astrophysique - CEA/DSM - CNRS - Universit\'{e} Paris Diderot, B\^{a}t. 709, CEA-Saclay, F-91191 Gif-sur-Yvette Cedex, France\\
\and
Laboratoire Traitement et Communication de l'Information, CNRS (UMR 5141) and T\'{e}l\'{e}com ParisTech, 46 rue Barrault F-75634 Paris Cedex 13, France\\
\and
Laboratoire de Physique Subatomique et de Cosmologie, Universit\'{e} Joseph Fourier Grenoble I, CNRS/IN2P3, Institut National Polytechnique de Grenoble, 53 rue des Martyrs, 38026 Grenoble cedex, France\\
\and
Laboratoire de Physique Th\'{e}orique, Universit\'{e} Paris-Sud 11 \& CNRS, B\^{a}timent 210, 91405 Orsay, France\\
\and
Lawrence Berkeley National Laboratory, Berkeley, California, U.S.A.\\
\and
Max-Planck-Institut f\"{u}r Astrophysik, Karl-Schwarzschild-Str. 1, 85741 Garching, Germany\\
\and
National University of Ireland, Department of Experimental Physics, Maynooth, Co. Kildare, Ireland\\
\and
Niels Bohr Institute, Blegdamsvej 17, Copenhagen, Denmark\\
\and
Optical Science Laboratory, University College London, Gower Street, London, U.K.\\
\and
SISSA, Astrophysics Sector, via Bonomea 265, 34136, Trieste, Italy\\
\and
SUPA, Institute for Astronomy, University of Edinburgh, Royal Observatory, Blackford Hill, Edinburgh EH9 3HJ, U.K.\\
\and
School of Physics and Astronomy, Cardiff University, Queens Buildings, The Parade, Cardiff, CF24 3AA, U.K.\\
\and
UPMC Univ Paris 06, UMR7095, 98 bis Boulevard Arago, F-75014, Paris, France\\
\and
Universit\'{e} de Toulouse, UPS-OMP, IRAP, F-31028 Toulouse cedex 4, France\\
\and
University of Granada, Departamento de F\'{\i}sica Te\'{o}rica y del Cosmos, Facultad de Ciencias, Granada, Spain\\
% ...
}

%% file: abstract.tex
\abstract{ {\bf Abstract:} 
We explore the 2013 \planck\ likelihood function with a high-precision
multi-dimensional minimizer (\minuit). This allows a
 refinement of the \lcdm\ best-fit solution with respect to
 previously-released results, and the construction of frequentist
 confidence intervals using profile likelihoods.
 The agreement with the cosmological results from the Bayesian framework is excellent,
 demonstrating the robustness of the \planck\ results to the statistical methodology.
 We investigate the inclusion of neutrino masses,
 where more significant differences may appear due to the
 non-Gaussian nature of the posterior mass distribution.
 By applying the Feldman--Cousins prescription, we again obtain
 results very similar to those of the Bayesian methodology.
 However, the profile-likelihood analysis of the CMB
 combination (\planck+WP+highL) reveals a minimum well within the
 unphysical negative-mass region.
 We show that inclusion of the \planck\ CMB-lensing information regularizes
 this issue, and provide a robust frequentist
 upper limit $\mnu \le 0.26\ev$ (95\% confidence) from the
 CMB+lensing+BAO data combination.
}


%% file: Tables/CLASS_setup.tex
\begingroup
\openup 5pt
\newdimen\tblskip \tblskip=5pt
\nointerlineskip
\vskip -3mm
\footnotesize
\setbox\tablebox=\vbox{
    \newdimen\digitwidth
    \setbox0=\hbox{\rm 0}
    \digitwidth=\wd0
    \catcode`"=\active
    \def"{\kern\digitwidth}
    \newdimen\signwidth
    \setbox0=\hbox{+}
    \signwidth=\wd0
    \catcode`!=\active
    \def!{\kern\signwidth}
\halign{
#\hfil\tabskip=1.5em&\hfil$#$\hfil\tabskip=0pt\cr
\noalign{\doubleline}
\omit\hfil \class parameter\hfil&\omit\hfil Value\hfil\cr
\noalign{\vskip 3pt\hrule\vskip 5pt}
tol\_background\_integration&10^{-3}\cr
tol\_thermo\_integration&10^{-3}\cr
tol\_perturb\_integration&10^{-6}\cr
reionization\_optical\_depth\_tol&10^{-5}\cr
l\_logstep&1.08\cr
l\_linstep&25\cr
perturb\_sampling\_stepsize&0.04\cr
delta\_l\_max&800 \cr
\noalign{\vskip 5pt\hrule\vskip 3pt}
} 
} 
\endPlancktable
\endgroup

%% file: Tables/CLASSvsCAMB.tex

\begingroup
\openup 5pt
\newdimen\tblskip \tblskip=5pt
\nointerlineskip
\vskip -3mm
\footnotesize
\setbox\tablebox=\vbox{
    \newdimen\digitwidth
    \setbox0=\hbox{\rm 0}
    \digitwidth=\wd0
    \catcode`"=\active
    \def"{\kern\digitwidth}
    \newdimen\signwidth
    \setbox0=\hbox{+}
    \signwidth=\wd0
    \catcode`!=\active
    \def!{\kern\signwidth}
\halign{
#\hfil\tabskip=1.5em&
\hfil$#$\hfil&
\hfil$#$\hfil\tabskip=0pt\cr
\noalign{\doubleline}
\omit\hfil Data set\hfil&\omit\hfil \camb \hfil&\omit\hfil \class\hfil\cr
\noalign{\vskip 3pt\hrule\vskip 5pt}
\cmb&10509.6&10509.9\cr
\cmbao&10510.8&10511.0\cr
\noalign{\vskip 5pt\hrule\vskip 3pt}
} 
} 
\endPlancktable
\endgroup

%% file: Tables/LCDM_bestfit.tex
\begingroup
\openup 5pt
\newdimen\tblskip \tblskip=5pt
\nointerlineskip
\vskip -3mm
\footnotesize
\setbox\tablebox=\vbox{
    \newdimen\digitwidth
    \setbox0=\hbox{\rm 0}
    \digitwidth=\wd0
    \catcode`*=\active
    \def*{\kern\digitwidth}
    \newdimen\signwidth
    \setbox0=\hbox{+}
    \signwidth=\wd0
    \catcode`!=\active
    \def!{\kern\signwidth}
    \newdimen\dotwidth
    \setbox0=\hbox{.}
    \dotwidth=\wd0
    \catcode`?=\active
    \def?{\kern\dotwidth}
\halign{
\hbox to 0.7in{$#$\leaderfil}\tabskip=0pt&
\hfil$#$\hfil&
\hfil$#$\hfil&
\hfil$#$\hfil&
\hfil$#$\hfil\tabskip=0pt\cr
\noalign{\doubleline}
\multispan1\hfil \hfil&\multispan2\hfil \cmb\hfil&\multispan2\hfil \cmbao\hfil\cr
\noalign{\vskip -3pt}
\omit&\multispan2\hrulefill&\multispan2\hrulefill\cr
\omit\hfil Parameter\hfil&\omit\hfil \COSMOMC\hfil&\omit\hfil \MINUIT\hfil&\omit\hfil \COSMOMC\hfil&\omit\hfil \MINUIT\hfil\cr
\noalign{\vskip 3pt\hrule\vskip 5pt}
H_0&*67.15**&*67.28**&*67.77**&*67.71**\cr
100\Omega_\mathrm{b}h^2&**2.207*&**2.210*&**2.216*&**2.216*\cr
\Omega_\mathrm{c}h^2&**0.1203&**0.1200&**0.1189&**0.1190\cr
n_\mathrm{s}&**0.9582&**0.9576&**0.9611&**0.9600\cr
\lnAs&**3.096*&**3.087*&**3.097*&**3.090*\cr
\zreio&*11.37**&*11.04**&*11.52**&*11.26**\cr
\noalign{\vskip 5pt\hrule\vskip 3pt}
A^\mathrm{PS}_\mathrm{100}&209?****&207?****&204?****&205?****\cr
A^\mathrm{PS}_\mathrm{143}&*72.6***&*73.5***&*71.8***&*73.0***\cr
A^\mathrm{PS}_\mathrm{217}&*59.5***&*61.1***&*59.4***&*60.7***\cr
A^\mathrm{CIB}_\mathrm{143}&**3.57**&**3.03**&**3.30**&**3.06**\cr
A^\mathrm{CIB}_\mathrm{217}&*53.9***&*51.2***&*53.0***&*51.2***\cr
A^\mathrm{tSZ}_\mathrm{143}&**5.17**&**4.00**&**4.86**&**4.01**\cr
r^\mathrm{PS}_\mathrm{143\times217}&**0.825*&**0.815*&**0.824*&**0.814*\cr
r^\mathrm{CIB}_\mathrm{143\times217}&**1.****&**1.****&**1.****&**1.****\cr
\gamma^\mathrm{CIB}&**0.674*&**0.647*&**0.667*&**0.647*\cr
c_\mathrm{100}&**1.****&**1.****&**1.****&**1.****\cr
c_\mathrm{217}&**0.997*&**0.997*&**0.997*&**0.997*\cr
\xi^\mathrm{tSZ-CIB}&**0.000*&**0.049*&**0.000*&**0.055*\cr
A^\mathrm{kSZ}&**0.89**&**2.87**&**1.58**&**2.89**\cr
\beta^1_1&**0.56**&**0.41**&**0.46**&**0.38**\cr
A^\mathrm{PS,ACT}_\mathrm{148}&*10.2***&*10.4***&*10.2***&*10.4***\cr
A^\mathrm{PS,ACT}_\mathrm{218}&*75.2***&*76.5***&*75.6***&*76.6***\cr
A^\mathrm{PS,SPT}_\mathrm{95}&**7.02**&**7.49**&**7.14**&**7.47**\cr
A^\mathrm{PS,SPT}_\mathrm{150}&**9.66**&**9.90**&**9.76**&**9.92**\cr
A^\mathrm{PS,SPT}_\mathrm{220}&*72.0***&*73.5***&*72.6***&*73.6***\cr
r^\mathrm{PS}_\mathrm{95\times150}&**0.830*&**0.787*&**0.806*&**0.790*\cr
r^\mathrm{PS}_\mathrm{95\times220}&**0.583*&**0.545*&**0.563*&**0.549*\cr
r^\mathrm{PS}_\mathrm{150\times220}&**0.908*&**0.915*&**0.911*&**0.915*\cr
A^\mathrm{ACTs}_\mathrm{dust}&**0.429*&**0.426*&**0.429*&**0.426*\cr
A^\mathrm{ACTe}_\mathrm{dust}&**0.879*&**0.845*&**0.843*&**0.844*\cr
y_\mathrm{ACTs}^\mathrm{148}&**0.991*&**0.991*&**0.992*&**0.991*\cr
y_\mathrm{ACTs}^\mathrm{217}&**1.****&**1.****&**1.****&**1.****\cr
y_\mathrm{ACTe}^\mathrm{148}&**0.987*&**0.987*&**0.988*&**0.988*\cr
y_\mathrm{ACTe}^\mathrm{217}&**0.960*&**0.961*&**0.961*&**0.962*\cr
y_\mathrm{SPT}^\mathrm{95}&**0.985*&**0.983*&**0.985*&**0.983*\cr
y_\mathrm{SPT}^\mathrm{150}&**0.984*&**0.984*&**0.985*&**0.985*\cr
y_\mathrm{SPT}^\mathrm{220}&**1.02**&**1.02**&**1.02**&**1.02**\cr
\noalign{\vskip 5pt\hrule\vskip 3pt}
\chi^2_\mathrm{min}      &10509.9*****        &10508.9*****        &10511.0*****       &10510.3*****\cr
\noalign{\vskip 5pt\hrule\vskip 3pt}
} 
} 
\endPlancktable
\endgroup

%% file: Tables/LCDM_profile.tex
\begingroup
\openup 5pt
\newdimen\tblskip \tblskip=5pt
\nointerlineskip
\vskip -3mm
\setbox\tablebox=\vbox{
    \newdimen\digitwidth
    \setbox0=\hbox{\rm 0}
    \digitwidth=\wd0
    \catcode`*=\active
    \def*{\kern\digitwidth}
    \newdimen\signwidth
    \setbox0=\hbox{+}
    \signwidth=\wd0
    \catcode`!=\active
    \def!{\kern\signwidth}
\halign{
\hbox to 0.7in{$#$\leaderfil}\tabskip=1.5em&
\hfil$#$\hfil&
\hfil$#$\hfil&
\hfil$#$\hfil&
\hfil$#$\hfil\tabskip=0pt\cr
\noalign{\doubleline}
\multispan1\hfil \hfil&\multispan2\hfil \cmb\hfil&\multispan2\hfil \cmbao\hfil\cr
\noalign{\vskip -3pt}
\omit&\multispan2\hrulefill&\multispan2\hrulefill\cr
\omit\hfil Parameter\hfil&\omit\hfil MCMC\hfil&\omit\hfil Profile-likelihood\hfil&\omit\hfil MCMC\hfil&\omit\hfil Profile-likelihood\hfil\cr
\noalign{\vskip 3pt\hrule\vskip 5pt}
\hz&67.3***\pm1.2***&67.2***\pm1.2***&67.8***\pm 0.8***&67.7***\pm0.8***\cr
100\omb&*2.207*\pm0.027*&*2.208*\pm0.027*&*2.214*\pm0.024*&*2.215*\pm0.024*\cr
\omc&*0.1198\pm0.0026&*0.1201\pm0.0026&*0.1187\pm0.0017&*0.1190\pm0.0017\cr
\ns&*0.9585\pm0.0070&*0.9575\pm0.0071&*0.9608\pm0.0054&*0.9598\pm0.0055\cr
\lnAs&*3.090*\pm0.025*&*3.087*\pm0.025*&*3.091*\pm0.025*&*3.088*\pm0.025*\cr
\zreio&11.2***\pm1.1***&11.0***\pm1.1***&11.2***\pm1.1***&11.2***\pm1.1***\cr 
\noalign{\vskip 5pt\hrule\vskip 3pt}
} 
} 
\endPlancktable
\endgroup

%% file: Tables/MNU_fit.tex
\begingroup
\openup 5pt
\newdimen\tblskip \tblskip=5pt
\nointerlineskip
\vskip -3mm
\footnotesize
\setbox\tablebox=\vbox{
    \newdimen\digitwidth
    \setbox0=\hbox{\rm 0}
    \digitwidth=\wd0
    \catcode`*=\active
    \def*{\kern\digitwidth}
    \newdimen\signwidth
    \setbox0=\hbox{+}
    \signwidth=\wd0
    \catcode`!=\active
    \def!{\kern\signwidth}
\halign{
#\hfil\tabskip=1.5em&
\hfil$#$\hfil&
\hfil$#$\hfil&
\hfil$#$\hfil\tabskip=0pt\cr
\noalign{\doubleline}
\omit\hfil Data set\hfil&\omit\hfil Fitted range\hfil&\omit\hfil $m_0$\hfil& \omit\hfil $\sigma_\nu$\hfil\cr
\noalign{\vskip 3pt\hrule\vskip 5pt}
CMB+lensing&[0,0.8]&!0.06&0.42\cr
CMB+lensing+BAO&[0,0.3]& -0.05 & 0.15\cr
\noalign{\vskip 5pt\hrule\vskip 3pt}
} 
} 
\endPlancktable
\endgroup

%% file: Tables/MNU_result.tex
\begingroup
\openup 5pt
\newdimen\tblskip \tblskip=5pt
\nointerlineskip
\vskip -3mm
\footnotesize
\setbox\tablebox=\vbox{
    \newdimen\digitwidth
    \setbox0=\hbox{\rm 0}
    \digitwidth=\wd0
    \catcode`"=\active
    \def"{\kern\digitwidth}
    \newdimen\signwidth
    \setbox0=\hbox{+}
    \signwidth=\wd0
    \catcode`!=\active
    \def!{\kern\signwidth}
\halign{
#\hfil\tabskip=1.5em&
\hfil$#$\hfil&
\hfil$#$\hfil\tabskip=0pt\cr
\noalign{\doubleline}
\omit\hfil Data set\hfil&\omit\hfil Bayesian posterior\hfil&\omit\hfil profile likelihood\hfil\cr
\noalign{\vskip 3pt\hrule\vskip 5pt}
CMB+lensing&0.85&0.88\cr
CMB+lensing+BAO&0.25&0.26 \cr
\noalign{\vskip 5pt\hrule\vskip 3pt}
} 
} 
\endPlancktable
\endgroup

%% file: acknowledgements.tex
\begin{acknowledgements}
We thank F. Le Diberder for discussions about the Feldman--Cousins method.
We gratefully acknowledge IN2P3 Computer Center
(\url{http://cc.in2p3.fr}) for providing the computing resources and
services needed to this work.
The development of \Planck\ has been supported by: ESA; CNES and CNRS/INSU-IN2P3-INP (France); ASI, CNR, and INAF (Italy); NASA and DoE (USA); STFC and UKSA (UK); CSIC, MICINN and JA (Spain); Tekes, AoF and CSC (Finland); DLR and MPG (Germany); CSA (Canada); DTU Space (Denmark); SER/SSO (Switzerland); RCN (Norway); SFI (Ireland); FCT/MCTES (Portugal); and PRACE (EU). A description of the \Planck\ Collaboration and a list of its members, including the technical or scientific activities in which they have been involved, can be found at \url{http://www.sciops.esa.int/index.php?project=planck&page=Planck_Collaboration}.
\end{acknowledgements}

%% file: references.tex
\bibliography{Planck_bib,refs}{}
\bibliographystyle{aa_arxiv}